Main Manuscript for

# Relativistic quantum decryption of large-scale neural coding

Sofia Karamintziou[1-6]                                              December 11, 2023

Research conducted, with personal funds, in the framework of the BridgeUSA Program with [1]Georgia Institute of Technology/School of Aerospace Engineering (2019-2022) [2]University of California, Irvine/Dept. of Mechanical and Aerospace Engineering (2018-2019) [3]Harvard Medical School/Massachusetts General Hospital (2017-2018) [4]University of California, Riverside/Dept. of Mechanical Engineering (2016-2017) and finalized at [5]NOVA University Lisbon/Dept. of Mathematics (2022)

[6]Current address: Centre for Research & Technology-Hellas/Information Technologies Institute, 6th km Harilaou-Thermi, 57001 Thessaloniki, Greece

Email: skaramintziou3@gatech.edu or skaramintziou@iti.gr

**Spin-geometrical projections, from the study of the human universe onto the study of the self-organizing brain, are herein leveraged to address certain concerns raised in latest neuroscience research, namely (i) the extent to which neural codes are multidimensional; (ii) the functional role of neural dark matter; (iii) the challenge to classical model frameworks posed by the needs for accurate interpretation of large-scale neural recordings linking brain and behavior. On the grounds of (hyper-)self-duality under (hyper-)mirror supersymmetry, relativistic quantum principles are introduced, whose consolidation, as pillars of a graphical game-theoretical construction, is conducive to (i) the high-precision reproduction and reinterpretation of core experimental observations on neural coding in the self-organizing brain, with the instantaneous geometric dimensionality of neural representations of a spontaneous behavioral state being proven to be *at most* 16, unidirectionally; (ii) the coexistability of ordinary and 'dark' neural coding for (co-)behavior; (iii) a possible role for spinor (co-)representations, as the latent building blocks of self-organizing cortical circuits subserving (co-)behavioral states; (iv) an early crystallization of pertinent multidimensional synaptic (co-)architectures, whereby Lorentz (co-)partitions are in principle verifiable; (v) the allusion to octonionic dynamics sustained by triality, as candidate dynamics underlying internal states or emotions; and, ultimately, (vi) potentially inverse insights into matter-antimatter asymmetry. New avenues for the decryption of large-scale neural coding in health and disease are being discussed.**

## Significance Statement

The instantaneous geometric dimensionality of neural representations of a spontaneous behavioral state has been experimentally documented to saturate at the value of $16$, unidirectionally. Here we provide computational evidence in support of this observation. We show that this phenomenon is mediated by a principle for (in)stability of a broad class of time-invariant, multidimensional self-organizing networked systems. In the self-organizing brain, such (non-)equilibrium state (W-duality or mirror supersymmetry) is sustained by a latent asymmetry: deep neural matter-antimatter asymmetry or a higher amount of excitatory vs. inhibitory neural deep clusters. Mirror supersymmetry is proven to break down at geometric dimensionality $2^q$, with $q = 5$. The reported spin-geometrical quantum principle unearths early neuronal crystals of behavior.

We consider the call[1] for new insights holding the potential to subserve an intensive and accurate interpretation of large-scale neural recordings linking brain and behavior. One has traditionally elaborated on neural coding and its relation to (sub)cortical circuit geometry by arguing, unilaterally, for the visible, ordinary scenario of correlated variability, as a signature of ongoing activity[2,3], and for either a low[4] or high[5,6] geometric dimensionality of neural representations. To argue in a manifold and balanced way, we recur to the tenets of spin geometry[7-11]. While it has been claimed[9,12-15] that these are central to the realms of special relativity, quantum computing and supersymmetric theories, we wish to learn more about their potential translation to our understanding of human brain function, at a level of population dynamics that does not obscure single-neuron interactions[1]. Conversely, by a technical treatment applicable to the field of neuroscience and beyond (TM), a link between instantiations of self-duality, mirror supersymmetry and relativistic quantum theory is principally drawn, adding new dimensions to previous observations[16,17].

We are particularly interested in self-organization, namely in spontaneous rather than evoked dynamics. In this respect, emphasis is being placed on the functional role of inhibition[18-21] in deeply clustered networks, innately suggestive of a super(a)symmetric non-cooperative game[22,23]. But, by U-duality under hyper-mirror super(a)symmetry – insinuating relativistic quantum entanglement[24,25] (Fig. 1b; Methods) – we maintain the coexistability of the aforesaid, ordinary neural coding scenario and its 'dark' counterpart[1,26]: a scenario of sparse coding in self-organizing neuronal circuits. And, by W-duality under mirror supersymmetry – essentially implying relativistic quantum superposition[27] (Fig. 1a; Methods) – we point to a bounded (Fig. 1c) rather than an arbitrarily high geometric dimensionality of neural representations of behavioral states, which, in light of experimental observations, suggests a distinction between two, possibly interacting, core processes: the 'economy' of spontaneous behaviors vs. a potential 'profligacy' of emotions. In the same vein, a super-space structure of the even Clifford algebra (Figs. 2-3) – conditionally intimating matter-antimatter symmetry violation[28] – is conducive to the high-precision reproduction and reinterpretation of core experimental observations on neural coding in the self-organizing brain, in connection with behavior and geometric dimensionality.

Spinor (co-)representations are emerging as the latent building blocks of neural modules sustained by W-duality, the spontaneous dynamics being attributable to mirror-symmetric (co-)flows. The proposed approach to disentanglement from complexity is proven to respect classical frameworks yet, at the same time, to implicitly address the inadequacy of their concepts and the challenges posed to them by the insights being offered by emerging technologies used to probe large-scale brain activity. Keeping pace with the concurrent development of quantum artificial intelligence[15,29-32], it is ultimately conceived to shed light on early keystone (co-)architectures of multidimensional synaptic economics (Figs. 4-5), whose alteration may be anticipated to bear significant correlation with pathological conditions (Table 1).

## I. At most 16: the economical correlates of spontaneous behavior under mirror supersymmetry

The super-space structure of the even Clifford algebra introduced in the accompanying material is herein conjectured to be conditionally interrelated with models of matter-antimatter asymmetry in the observable universe[28]. The hypothesis complements earlier assertions[14] and is substantiated both theoretically and experimentally below, at the level of deep neural matter-antimatter asymmetry. More specifically, the structure is exploited to implicitly distinguish between excitatory vs. inhibitory neural populations and, by reduction, between neural matter vs. neural antimatter in self-organizing (sub)cortical networks (Table 1): that is, we define *neural antimatter*† and, in particular, neural antiparticles, as the particles of inhibitory neurons.

We henceforth let a self-organizing (sub)cortical circuit be denoted by $\mathscr{O}$. Spinor (co-)representations give rise to rotational/Lorentzian (co-)dynamics. Such (co-)dynamics have, inter alia, been observed in the cortex[33,34]. Accordingly, we will broadly distinguish between two neural coding scenarios, namely

(i) chaotic correlations, as a reduced form of correlated variability (also termed shared variability)[2,3,35]. For us, such correlations originate in a hybrid (single-neuron/population-level) structure – the spin-geometrical flow network $\mathscr{F}$ underlying $\mathscr{O}$, on which a dilemma game occurs in reduced form (TM);

(ii) chaotic co-correlations, perceived as the 'dark' or silent[1,26] counterpart of chaotic correlations, i.e., as a reduced form of co-correlated variability. A scenario of sparse coding in self-organizing neuronal circuits is of relevance here. Accordingly, we shall define *neural dark matter*, as the particles of neurons in $\mathscr{O}$ that are sparsely active. Chaotic co-correlations are being considered to originate in the co-flow network $\mathscr{F}^{\perp}$ underlying $\mathscr{O}$, on which a co-dilemma game occurs in reduced form.

---

†Note that neural antimatter is not necessarily perceived as antimatter in the conventional sense.

By virtue of principle 1 in Methods:

α. A self-organizing, supersymmetric (sub)cortical circuit, $\mathscr{C}$ , whose latent building block is an irreducible representation of $\mathbf{Spin}^+\left(1,5\right)$, exhibits Nash-Simon dimensionality of neural representations of a behavioral state $16$ or $0$, in the same direction of flow (at most $32$ , in opposite directions of flow).

β. A self-organizing, supersymmetric (sub)cortical circuit, $\mathscr{C}$ , whose latent building block is an irreducible co-representation of $\mathbf{Spin}^+\left(1,5\right)$, exhibits Nash-Simon dimensionality of neural co-representations of a co-behavioral state $16$ or $0$, in the same direction of co-flow (at most $32$ , in opposite directions of co-flow).

We may translate these statements.

a. Under the visible, ordinary code of chaotic correlations (correlated variability), a self-organizing (sub)cortical circuit, $\mathscr{C}$ , with orientation-preserving, Lorentzian dynamics, exhibits instantaneous geometric dimensionality of neural representations of a behavioral state at most $16$ , in the same direction of flow (at most $32$ , in opposite directions of flow).

b. Under the 'dark' code of chaotic co-correlations (co-correlated variability), a self-organizing (sub)cortical circuit, $\mathscr{C}$ , with orientation-preserving, Lorentzian co-dynamics, exhibits instantaneous geometric dimensionality of neural co-representations of a co-behavioral state at most $16$ , in the same direction of co-flow (at most $32$ , in opposite directions of co-flow).

**Experimental validation.** Claim a reproduces the precise number of dimensions of behavioral videographic information, at which predictability of spontaneous cortical, as well as brain-wide activity in awake mice turns out to saturate[5] under the coding scenario of correlated variability, optimal predictability of neural activity being linked to instantaneous behavior. In this experiment, neural population activity modulated almost all neurons in the same direction.

**Candidate dynamics for internal states or emotions: re-interpreting experimental data.** The outcome of the aforementioned experiment, with regard to (geometric) dimensionality, is misconstrued in the respective study[5]. Strong emphasis is placed on a statistical analysis – termed shared-variance component analysis – implemented to identify potentially additional latent dimensions encoded by spontaneous cortical activity, with reference to the animal's behavioral state, as assessed via orofacial movements. A high dimensionality ($=128$ ) is brought up as an outcome of the analysis, under the coding scenario of correlated variability. Given this outcome, the videographically assessed dimensionality ($=16$), despite being a saturation value, is presented as a lower bound of a sequence of increasing dimensionality values ("visual cortex encodes at least $16$ dimensions of motor information"). The discrepancy between the outcome of the analysis and that of the videography is attributed to motor activity not visible on the face and, as such, decodable only by advanced methods, or, to purely internal cognitive states underlying behavior.

Let us assume that the unidentified (non-arousal) state indeed reflects motor information. The dynamics allied to this state would then be sustained by either real or complex spinor representations. The latter case does not yield mirror supersymmetry. The former case would fall within a generalization of claim a from Lorentzian to real spinor representations, yielding the same saturation value ($=16$).

It, thereby, turns out that the dynamics pertinent to the unidentified state are distinct from the dynamics pertinent to motor information and reflect an exceptional phenomenon representable at exactly $128$ dimensions and at no other dimension. It is reservedly conjectured that such dynamics are octonionic and are sustained by an esoteric structure − a triality[9,12], which comprises a restriction of a polynomial homomorphism of the $128$-dimensional even Clifford algebra $\mathfrak{C}_{0,8}^0 \supset \mathbf{Spin}\left(0,8\right)$. Such phenomena may underly the neural representation of certain internal states[36] or emotions[37].

Concisely, we maintain that, in the same direction of flow, there are no ongoing behavioral states between the $16$-dimensional (real spinorial) dynamics and the $128$-dimensional (possibly octonionic) dynamics. Instead, a distinction has to be drawn between a process that imposes an upper bound ($=16$) on dimensionality and a process that takes place at a specific dimension ($=128$). At the same time, we have currently no reason to consider it unlikely that the two processes interact, in which case the partial predictability of the unidentified latent state by the mouse's ongoing behavior would as well be explainable. And so would be the fact that facial movements are indicators of emotions[37]. As a matter of fact, while we point to two distinct core processes regulating the geometric dimensionality of neural representations in self-organizing neural circuits, we defy a classification in a sterile dichotomy.

Here are some further remarks in support of the above claims. Exactly $32$ dimensions of natural images (the natural analogue of drifting gratings spanning opposite directions), have been linked to increased accuracy in the estimation of trial-average stimulus responses in the visual cortex of awake mice[6,38]. In these particular experiments, stimulus responses were sparse. To the exclusion of cognitive and behavioral state-variables, a stimulus-related dimensionality value equaling $2800$ has been reported[6], as assessed in the main experimental phase. We found no evidence for saturation at any dimensionality higher than $32$. Texture representation in primate somatosensory cortex can be $21$-dimensional across opposite features of texture (fine vs. coarse)[39]. The use of mesoscale calcium imaging to record cortex-wide neural activity in awake mice[4], under multiple behavioral environments, yielded coarsely $14$-dimensional spatiotemporal dynamics, over opposite directions of flow.

The low dimensionality ($=1$) of neural representations of arousal (pupil diameter) has been perceived[5] as being part of the dimensionality of neural representations of locomotion ($<16$), which, in turn, has been perceived as being part of the dimensionality of neural representations of orofacial movements ($=16$). We here point out that the dynamics allied to arousal are distinct from the dynamics allied to locomotion and orofacial movements, the former being rotational, the latter non-rotational. This remark is in agreement with a previous study[40]. By the same token, super-asymmetry, in the context of the present study, is distinct from supersymmetry. Recursively, two broad behavioral states previously observed[4] reflect the fact that the dynamics allied to locomotion vs. facial movements are sustained by distinct real spinor representations. A further distinction between the dynamics subserving different kinds of locomotion is possible. And, ultimately, interactions between the aforementioned (distinct) processes should be feasible, effectuating functional flexibility.

The following statement offers an additional explanation for the bound in claim a (Figs. 1-2).

**Theorem.** *The instantaneous geometric dimensionality of neural representations of a behavioral state, in a self-organizing, supersymmetric (sub)cortical circuit, $\mathscr{C}$, is restrained by deep neural matter-antimatter asymmetry.*

*Proof.* Assume that $\mathscr{C}$ exhibits instantaneous geometric dimensionality of neural representations of a behavioral state $d_i = 32$, in the same direction of flow. Then, by the statements in the accompanying material, deep neural matter-antimatter asymmetry in $\mathscr{C}$ vanishes non-trivially, which is a contradiction. The contradiction does not occur for $2 \leq d_i \leq 16$. ∎

## II. Coexistability: U-duals of W-duals

Principle 2 in Methods suggests (see also TM: III.5, III.6 and IV.5, IV.6):

**1.** In a self-organizing (sub)cortical circuit, $\mathscr{C}$, chaotic correlations sustained by rotational/Lorentzian dynamics and chaotic co-correlations sustained by rotational/Lorentzian co-dynamics are coexistable, the two neural codes being U-dual to each other and each exhibiting W-duality under mirror super(a)symmetry.

Thence, whenever they coexist, the two neural codes may be viewed as the two distinct 'sides of the same coin', each side exhibiting itself self-duality (Fig. 1b); in particular, W-duality under mirror super(a)symmetry. If we further account for opposite directions of flow, we obtain a pair of 'opposing' W-dualities on every side of the coin.

**2.**1. A self-organizing, supersymmetric (sub)cortical circuit, $\mathscr{C}$, whose latent building blocks are irreducible (co-)representations of $\mathbf{Spin}^+(1,5)$, exhibits complete Nash-Simon dimensionality of neural (co-)representations of (co-)behavioral states at most $32$, in the same direction of (co-)flow (at most $64$, in opposite directions of (co-)flow).

We may rephrase this statement (for an explanation, see Fig. 3 in addition to Methods).

**2.**2. Under the neural codes of chaotic (co-)correlations, a self-organizing, supersymmetric (sub)cortical circuit, $\mathscr{C}$, with orientation-preserving, Lorentzian (co-)dynamics, exhibits complete, instantaneous geometric dimensionality of neural (co-)representations of (co-)behavioral states at most $32$, in the same direction of (co-)flow (at most $64$, in opposite directions of (co-)flow).

**Experimental evidence.** While no experimental studies on the orthogonality of instantaneous multidimensional neural (co-)representations in self-organizing neuronal circuits have yet been conducted, the orthogonality of neural (co-)representations in the primary visual cortex during stimulus presentation has been addressed[5,41]. Neural (co-)representations of sensory inputs and behavioral states were found[5] to be mixed together in the same cell population.

The distinct nature of super-asymmetric neural circuits (TM: II.5) may best be reflected in the following.

**3**.1. A self-organizing, super-asymmetric (sub)cortical circuit, whose latent building blocks are

(i) real spinor representations, exhibits Nash-Simon dimensionality of neural representations of an arousal or behavioral state $1$ or $0$, in the same direction of flow.
(ii) real spinor co-representations, exhibits Nash-Simon dimensionality of neural co-representations of a co-arousal or co-behavioral state $1$ or $0$, in the same direction of co-flow.
(iii) real spinor (co-)representations, exhibits complete Nash-Simon dimensionality of neural (co-)representations of (co-)behavioral states at most $2$, in the same direction of (co-)flow.

Let us translate.

**3**.2. Under the neural code(s) of

(i) chaotic correlations, a self-organizing (sub)cortical circuit, with orientation-preserving, rotational dynamics, exhibits instantaneous geometric dimensionality of neural representations of an arousal or behavioral state at most $1$, in the same direction of flow.
(ii) chaotic co-correlations, a self-organizing (sub)cortical circuit, with orientation-preserving, rotational co-dynamics, exhibits instantaneous geometric dimensionality of neural co-representations of a co-arousal or co-behavioral state at most $1$, in the same direction of co-flow.
(iii) chaotic (co-)correlations, a self-organizing (sub)cortical circuit, with orientation-preserving, rotational (co-)dynamics, exhibits complete, instantaneous geometric dimensionality of neural (co-)representations of (co-)behavioral states at most $2$, in the same direction of (co-)flow.

**Experimental evidence.** Hyper-self-duality under rotational (co-)dynamics is reported in a visually guided reaching task[34], involving primates. During the task, the supplementary motor area (SMA) and motor cortex displayed similar single-neuron responses. However, while population dynamics dominated by rotations subserving muscle activity were prominent in the motor cortex, such dynamics were absent in the SMA, where activity patterns subserved movement initiation.

By now, it should be clear enough:

**4.** Super(a)symmetry in the flow network $\mathcal{F}$ underlying $\mathscr{O}$ respects the two major types of correlated variability observable in $\mathscr{O}$.

Aspects of the two types of correlated variability, arising through a distinction between microscopic (homogeneous) vs. macroscopic (heterogeneous) forms of neural activity, have been discussed both theoretically[35] and experimentally[42]. A subdivision by inhibitory neural type has further been attempted[3], while recurrent vs. feedforward structures are known to subserve distinct functions[43].

**5.** Under differential correlations, $\mathscr{O}$ exhibits instantaneous geometric dimensionality of neural representations $0$.

Indeed, differential correlations are information-limiting[44].

## III. Lorentz crystals

Implications of coexistability for the early crystallization of multidimensional synaptic (co-)architectures may best be alluded to by what follows (see also TM: III.2, III.3). Consider a self-organizing, supersymmetric neuronal circuit, $\mathscr{O}$, whose latent building block is an irreducible representation (co-representation) of $\mathbf{Spin}^{+}(1, q)$, with $2 \leq q \leq 4$. If $\mathscr{O}$ exhibits instantaneous geometric dimensionality of neural representations of a behavioral state (resp., neural co-representations of a co-behavioral state) $d_i = 2^q$, in the same direction of flow (resp., of co-flow), then:

(i) there is a unique strong (resp., co-strong) cluster in $\mathscr{O}$ — the source (resp., the co-source), whose neurons do not (resp., do) receive input from neurons not in the cluster, and which includes neurons from every neural subpopulation (i.e., every deep cluster) in $\mathscr{O}$;
(ii) the source's (resp., co-source's) geometry exhibits W-duality under mirror supersymmetry;
(iii) a bipartition (resp., co-bipartition) of $\mathscr{O}$'s geometrical source-minor (resp., co-source-minor) into two areas $\alpha$ and $\beta$ forms;
(iv) a Lorentz, $(1, q)$-partition (resp., co-$(1, q)$-partition) of $\mathscr{O}$'s geometrical source-minor (resp., co-source-minor) forms;

(v) a source's (resp. co-source's) neuron in a neural subpopulation located in area $\alpha$ in $\mathscr{O}$, receives input from neurons of at most (resp., at least) the same neural subpopulation in area $\alpha$ and of at most (resp., at least) one other neural subpopulation located in area $\beta$ in $\mathscr{O}$; $\alpha \neq \beta$;

or:

(vi) there is a unique strong (resp., co-strong) cluster in $\mathscr{O}$ — the basin (resp., the co-basin), whose neurons do not (resp., do) provide input to neurons not in the cluster, and which includes neurons from every neural subpopulation (i.e., every deep cluster) in $\mathscr{O}$;

(vii) the basin's (resp., co-basin's) geometry exhibits W-duality under mirror supersymmetry;

(viii) a bipartition (resp., co-bipartition) of $\mathscr{O}$'s geometrical basin-minor (resp., co-basin-minor) into two areas $\mathscr{k}$ and $\ell$ forms;

(ix) a Lorentz, $(1,q)$-partition (resp., co-$(1,q)$-partition) of $\mathscr{O}$'s geometrical basin-minor (resp., co-basin-minor) forms;

(x) a basin's (resp. co-basin's) neuron in a neural subpopulation located in area $\mathscr{k}$ in $\mathscr{O}$, provides input to neurons of at most (resp., at least) the same neural subpopulation in area $\mathscr{k}$ and of at most (resp., at least) one other neural subpopulation located in area $\ell$ in $\mathscr{O}$; $\mathscr{k} \neq \ell$.

Thence, what is commonly perceived as sparse activity of 'dark' neurons[1,26] in cortical circuits can in principle be sustained by a highly symmetric structure.

**Experimental evidence.** The inclination of cortical networks of pyramidal neurons and inhibitory interneurons to form multipartitions has already been rigorously reported[45]. Guiding this evidence further, we point to Lorentz (co-)crystals (Fig. 4(a)-(b); Fig. 5).

Let us note that co-partitions imply induced cliques (in particular, semicomplete subgraphs[46]; Fig. 4(b); Fig. 5(b)).

We leave open the question of whether neural mirror matter and associated structures[47] (Fig. 4(c)-(d)) are implicitly allied to cortical rhythmic activity during sleep (TM: IV.5).

## Discussion

A spin-geometrical construction has been unveiled, which, while conditionally intimating matter-antimatter asymmetry, has been leveraged to address certain challenges posed by the emergent needs to construe large-scale neural recordings linking brain and spontaneous behavior. Within a quantum computational framework, it is conducive to the reproduction, reinterpretation and guidance of experimental outcomes on neural coding in the self-organizing brain, in connection with behavior and geometric dimensionality; and, for that matter, to a possible role for spinor representations as the latent building blocks of self-organizing cortical circuits subserving behavioral states. The principles obtained suggest that (i) neural matter-antimatter asymmetry restrains the instantaneous geometric dimensionality of neural representations of a spontaneous behavioral state, this dimensionality being at most 16, in the same direction of flow; (ii) the ordinary and 'dark' neural coding scenarios are coexistable and sustainable by highly symmetric structures, dual to each other, with Lorentz (co-)crystals being characteristic of these structures.

The hypothesis that neural matter is the sole fundamental constituent of neural circuits would fail to provide a complete, biologically sensible explanation for the reported bound on dimensionality. Indeed, nature turns out to be asymmetrically symmetric. By ignoring inhibition – and neural antimatter as a reduction thereof – one ignores asymmetry, as much as balance and stability[48]. Two types of neural antimatter are further implicit to the distinction between adapting and fast-spiking interneurons[20], the latter being linked to synaptic reciprocity, macroscopic neural activity, supersymmetry and Lorentzian dynamics (TM: II.5).

Defying dogmatic and unilateral views, we may contemplate[49] the development of rigorous methods that effectively leverage insights from cosmic into brain-wide structures, namely from physics beyond the Standard Model into neuroscience beyond classical model frameworks and therefrom to the identification of key therapeutic targets. Towards an expanded model framework subserving large-scale neural data in health and disease, we suggest that inference of multidimensional synaptic (co-)architectures may be further facilitated by an extension of the existing model, based on (i) a dynamic neural matter-antimatter approach, as a test-bed for homeostatic plasticity mechanisms; (ii) forces - neural matter interactions, as a test-bed for stimulus responses; (iii) octonionic dynamics, as a test-bed for the dynamics underlying internal states or emotions; and, ultimately, (iv) broken mirror supersymmetry, as a test-bed for pathological deep neuronal network mechanisms.

In section I, we have already discussed clues supporting octonionic dynamics as candidate dynamics for internal states or emotions. Here, we briefly comment on the rest of the proposed future directions.

Homeostatic plasticity[43,50-52] is innately suggestive of a dynamic analogue of the presented neural matter - antimatter approach. In particular, the supergame of the dilemma game underlying $\mathscr{G}$, along with appropriately designed finite-memory models involving retaliation for violations, is a dynamic construction through which a wide spectrum of self-enforced efficient outcomes, otherwise observable in cooperative games (an analogue of evoked dynamics) may be attained[53]. Such approach may hold the potential to elucidate self-organizing, deep neuronal network (co-)architectures driven by homeostatic plasticity mechanisms and their relation to neural matter[54,55]. In addition, we speculate, geometrical features of short-term processes[56] may start becoming known to us by appropriately adjusting relevant treatments in automatic control theory[57,58]. Though these processes are of irrefutable significance, we did not herein delve into analyses of this kind.

It has been argued[14] that the supersymmetry spanned by bosons (displaying integer spin and carrying the forces) and fermions (displaying half-integer spin and comprising ordinary matter) is a reflection of and intimately related to the superalgebra structure of the Clifford algebra. In the language of theoretical physics and quantum mechanics, the interrelation of the Clifford algebra with the Schrödinger-Pauli equation, the Dirac equation, as well as Weyl, Majorana and Dirac spinors has been discussed by several authors[8-10]; in the language of quantum computing, Clifford gates[13] are elements of the Clifford-Lipschitz group, which lies in the multiplicative group of invertible elements of the Clifford algebra. Against the backdrop of this study, we claim that parallels between a spin-geometrical framework and evoked dynamics (stimulus responses) in neuronal circuits can be drawn, illuminating quantum network mechanisms in pertinent processes.

Drawing upon a clinical-data-driven approach[59], we would like to cautiously leverage the present observations on neural coding of motor information, as they pointed to normal neuronal network mechanisms, whose alteration, suggestive of broken supersymmetry, may bear significant correlation with the pathological underpinnings of amyotrophic lateral sclerosis (ALS); these are characterized by a profound synapse loss prior to the onset of symptomatic motor decline[60] and hyperexcitability[61] of the motor circuit. Outcomes in the context of the proposed dynamic neural matter-antimatter approach may be suggestive of further critical processes, since, based on ALS models, disease manifestations at the neuromuscular junction are dynamic[62]. Despite distinct biological underpinnings which are not to be ignored, epilepsy[63] is similarly characterized by the aberrant activation and hyperexcitability[64] of neuronal circuits; we are interested in pertinent deep neuronal network alterations, as well as alterations in the circuit geometry of homeostatic plasticity mechanisms potentially underlying the disorder[65].

Coherent energy transfer and quantum signaling across microtubule networks have been postulated to be compromised in neurodegenerative tauopathic disease, such as Alzheimer's disease (AD), due to ineffective channeling of photons for signaling or dissipation[66]. In patients with AD, default-mode network activity − i.e., ongoing metabolic activity in the resting brain during episodic and autobiographical memory retrieval or inwardly oriented mental activity − and, particularly, resting-state activity in the posterior cingulate cortex and hippocampus is significantly altered compared to healthy controls[67], suggesting disrupted connectivity between these two regions. Furthermore, failures in homeostatic synaptic plasticity − and, as such, in homeostatic regulation of excitatory-inhibitory balance − in cortico-hippocampal circuits have been proposed to represent the driving force of early disease progression[68]. On the basis of an innate projection of excitatory-inhibitory neural populations onto neural matter-antimatter, it remains an open question whether profoundly altered large-scale circuit geometry due to broken supersymmetry, and weak quantum effects due to deficient quantum signaling, during dynamic interactions between photons and neural fermions-antifermions, may underlie and explain pathological homeostatic synaptic plasticity mechanisms, as well as disrupted functional connectivity and organization across the aforesaid neurobiological structures, in AD.

Throughout man's never-ceasing endeavor to understand one[1,49,69,70] of the most convoluted systems in nature, including the human universe in itself[71], it is ultimately wise to be reminded of incompleteness, in Gödel's sense[72]. All the same, rather than denying[73] progress beyond classical models, we may bring to the light of day the foresight that quantum mind, brain and consciousness[74] are explorable without recurring to mysticism. Relativity is afoot, listening to the (loud) sound of silence.

## METHODS

### Relativistic quantum principles

**1.** The fist pillar of the present approach, derived by virtue of the tenets of spin geometry, is the following principle, establishing, in light of experimental observations[5], a bounded geometric dimensionality of neural representations of behavioral states (for notational and other technical details, we refer the reader to the accompanying material).

**Proposition.** *Let* $\vartheta : X \to \mathbb{R}$ *denote a nondegenerate quadratic form on an* $\mathbb{R}$ *-linear space* $X$ *, with* $\dim_{\mathbb{R}} X = p + q < \infty$ *; therewith,* $(X, \vartheta)$ *is isomorphic to the pseudo-Euclidean space* $\mathbb{R}^{p,q}$ *, with signature* $(p,q)$ *and positive index* $p$ *. In* $\dim_{\mathbb{R}} X = p + q \leq 5$ *,*

$$\mathcal{O}(\delta) = \mathcal{O}_{p,q}^{0} \, . \tag{1}$$

*That is, the null super-cone,* $\mathcal{O}(\delta)$ *, of the quadratic norm,* $\delta$ *, on the even Clifford algebra,* $\mathbb{C}_{p,q}^{0}$ *, of* $(X, \vartheta)$ *is self-dual to the common boundary,* $\mathcal{O}_{p,q}^{0}$ *, of the disjoint open sets* $\Gamma_{p,q}^{0+}$ *and* $\Gamma_{p,q}^{0} \setminus \Gamma_{p,q}^{0+}$ *,* $\Gamma_{p,q}^{0+}$ *denoting the identity component of the special Clifford-Lipschitz group* $\Gamma_{p,q}^{0}$ *. The statement is not true in* $\dim_{\mathbb{R}} X = p + q = 6$ *, with* $p \neq 0$ *, where mirror supersymmetry breaks down.*

The form of self-duality observed in the above proposition we characterize as Wittgenstein or W-duality under mirror super(a)symmetry. By (1), the respective spontaneous dynamics of $\mathcal{O}$ are attributable to relativistic quantum superposition (TM: II.1.3).

**2.** Given the mirror-super(a)symmetric flow and co-flow network, $\mathcal{F}_{*}$ and $\mathcal{F}_{*}^{\perp}$, conditionally underlying $\mathcal{O}$, hyper-self-duality under hyper-mirror super(a)symmetry,

$$\mathcal{F}_{*} = \mathcal{F}_{*}^{\perp} \, , \tag{2}$$

comprises the second pillar of the model framework, implying, in light of experimental observations[34], the coexistability of ordinary and 'dark' neural coding for (co-)behavior. This form of self-duality we characterize as unifying or U-duality. By (2), the respective spontaneous (co-)dynamics of $\mathcal{O}$ are attributable to relativistic quantum entanglement.

(a)

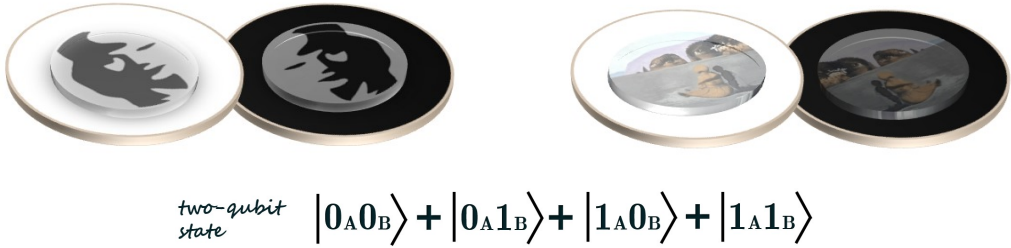

(b)

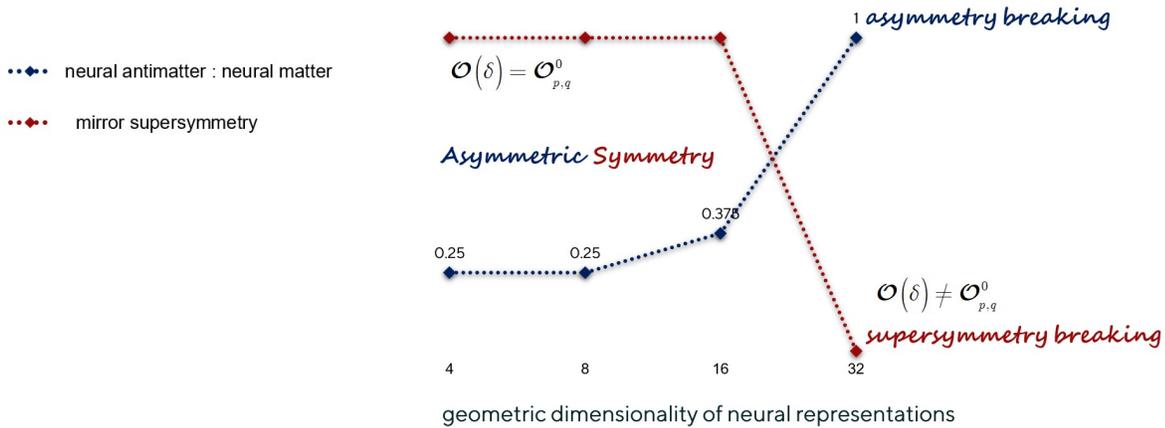

(c)

**Fig. 1. Graphical abstract.** (a) Illustrations of W-duality implying quantum superposition (for simplicity of illustration, states are not normalized to length 1). (b) Illustrations of U-duality implying quantum entanglement: (upper left) W-duality on each side (white and black) of the same coin; (upper right) In large-scale, self-organizing neural circuits, mirror supersymmetries are sustained by deep asymmetries, artistically analogous to the disproportion of colors underlying a painting [inner image: adapted from *Adolescence*, Salvador Dalí (1941)]. (c) W-duality under mirror supersymmetry is economically effectuated by a bound on geometric dimensionality: it is attainable at geometric dimensionality values at most 16 (unidirectionally) and is sustained by a latent asymmetry (deep neural matter-antimatter asymmetry), therewith essentially yielding asymmetrically symmetric states [for visualization purposes, we arbitrarily assign the value '1' to the existence of mirror supersymmetry and the value '0' to mirror supersymmetry breaking; the blue dotted line depicts the ratio of inhibitory to excitatory neural deep clusters as a function of the geometric dimensionality of neural representations of a behavioral state].

Table 1. From physics beyond the Standard Model into neuroscience beyond classical model frameworks, on the grounds of relativistic quantum principles.

| Physics beyond the Standard Model | Neuroscience beyond 'the Standard Model' | | | | |
|---|---|---|---|---|---|
| | Projections \| Neural Coding | | Implications for the understanding/assessment of | Pathological Mechanisms \| Pathology | |
| (ordinary) matter vs. dark matter | neural (ordinary) matter vs. neural dark matter | chaotic correlations vs. chaotic co-correlations | latent states (behavioral, cognitive, etc.) vs. latent co-states in wakefulness | Broken supersymmetry \| Altered neuronal network excitability and homeostatic plasticity \| Abnormal multidimensional (co-)architectures | Genetic or Acquired Neurodegenerative Diseases – Motor Neuron Disease \| Neuronal Excitability Disorders – Epilepsy \| Cognitive Impairment \| Mental Health Disorders |
| matter vs. antimatter | neural matter vs. neural antimatter | excitation vs. inhibition under chaotic correlations | multidimensional architectures of latent states in wakefulness | | |
| dark matter vs. dark antimatter | neural dark matter vs. neural dark antimatter | excitation vs. inhibition under chaotic co-correlations | multidimensional co-architectures of latent co-states in wakefulness | | |
| mirror matter vs. mirror dark matter | neural mirror matter vs. neural mirror dark matter | mirror chaotic correlations vs. mirror chaotic co-correlations | REM vs. NREM sleep (Hypothesis) | | |
| weak vs. strong interactions | unilateral vs. reciprocal connections | superasymmetry vs. supersymmetry | adapting interneurons vs. fast-spiking interneurons | | |
| photons vs. matter | stimuli vs. neural matter | chaotic (co-)correlations | stimulus responses | | |
| Pillars: Inter-relativistic Quantum Principles / (Hyper-)Self-Duality under (Hyper-)Mirror Supersymmetry | | | | | |

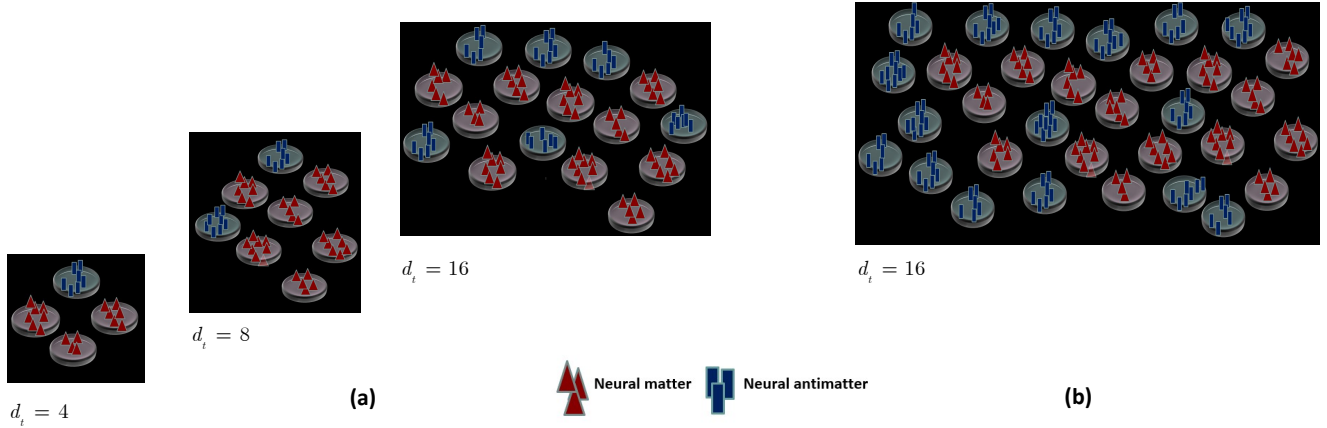

$d_i = 4$

$d_i = 8$

**(a)**

$d_i = 16$

$d_i = 16$

**(b)**

🔺 Neural matter ▮ Neural antimatter

**Fig. 2. An explanation for the bound in I.a and the spontaneous breaking of mirror supersymmetry.** Each button-shaped item in (a) represents a deep cluster (neural subpopulation at a distinct dimension) in three different spin-geometrical configurations of a self-organizing, mirror-supersymmetric (sub)cortical circuit, $\mathscr{O}$: these configurations correspond to the super-space structure of the $d_i$- dimensional even Clifford algebra (TM). Assume that $\mathscr{O}$ exhibits instantaneous geometric dimensionality of neural representations of a behavioral state $d_i = 32$, in the same direction of flow. Then, deep neural matter-antimatter asymmetry in $\mathscr{O}$ vanishes non-trivially (b), which is a contradiction. The contradiction does not occur for (a) $4 \leq d_i \leq 16$ (values are given in the same direction of flow). Mirror supersymmetry in (b) is in fact feasible by at most $16$ out of $32$ clusters, with at most $6$ deep clusters consisting of neural antimatter. Recall (TM) that the instantaneous geometric dimensionality exhibited by $\mathscr{O}$ can be either $0$ or of the form $d_i = 2^q$, $q \geq 1$ (deep neural matter-antimatter asymmetry in $\mathscr{O}$ vanishes trivially for $q = 1$).

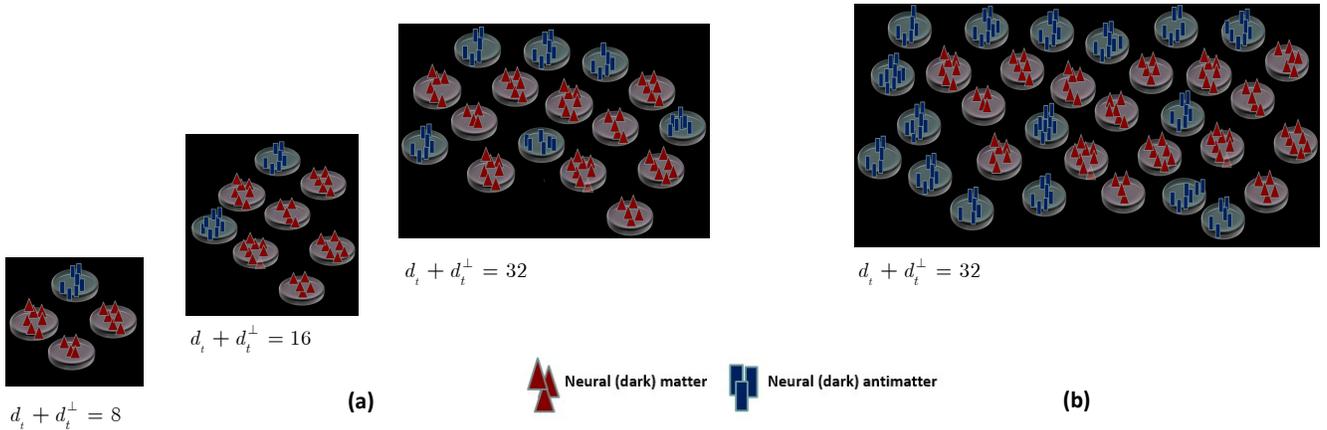

$d_i + d_i^\perp = 8$

$d_i + d_i^\perp = 16$

**(a)**

$d_i + d_i^\perp = 32$

$d_i + d_i^\perp = 32$

**(b)**

🔺 Neural (dark) matter ▮ Neural (dark) antimatter

**Fig. 3. A (co-)explanation for the bound in II.2.2 and the spontaneous breaking of hyper-mirror supersymmetry.** Each button-shaped item in (a) represents a deep cluster in three different spin-geometrical configurations of a self-organizing, hyper-mirror-supersymmetric (sub)cortical circuit, $\mathscr{O}$. Assume that $\mathscr{O}$ exhibits complete, instantaneous geometric dimensionality of neural (co-)representations of a (co-)behavioral state $d_i + d_i^\perp = 64$, in the same direction of (co-)flow. Then, deep neural (dark) matter-(dark) antimatter asymmetry in $\mathscr{O}$ vanishes non-trivially (b), which is a contradiction. The contradiction does not occur for (a) $8 \leq d_i + d_i^\perp \leq 32$ (values are given in the same direction of (co-)flow). Hyper-mirror supersymmetry in (b) is in fact feasible by at most $16$ out of $32$ clusters, with at most $6$ deep clusters consisting of neural (dark) antimatter.

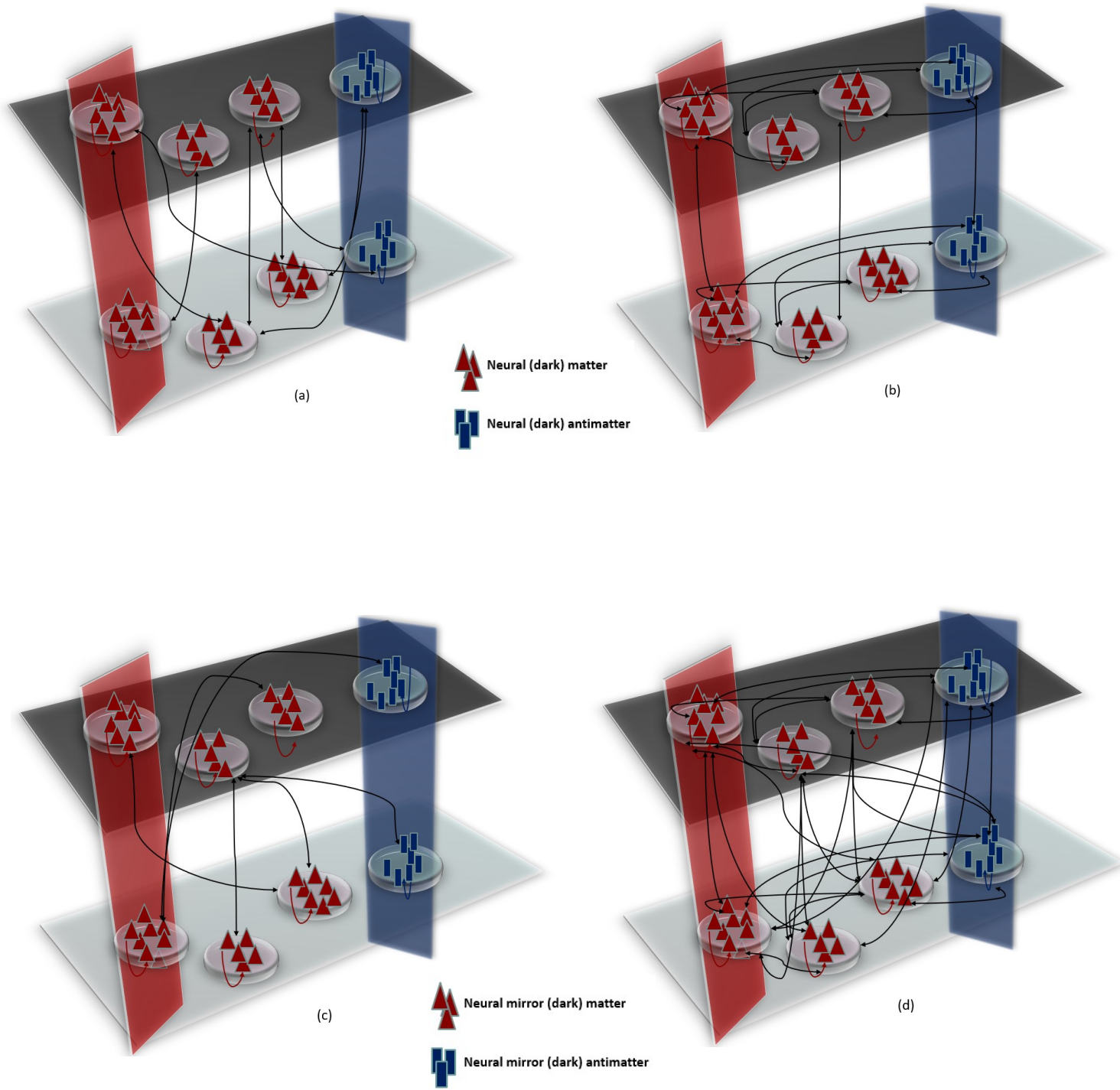

**Fig. 4. Lorentz 'spacetime' crystals.** (a) A Lorentz, $(1,3)$-partition ((b) A Lorentz, co-$(1,3)$-partition) on a bipartition (resp., co-bipartition) of $\mathscr{G}$ 's spin-geometrical minor: any two distinct deep clusters contained in same-colored layers are necessarily disconnected (resp., induce a semicomplete subdigraph); (c) The mirror Lorentz, $(1,3)$-partition of (a); (d) The mirror Lorentz, co-$(1,3)$-partition of (b). Note that $1$ out of $1+3$ layers consists exclusively of neural (dark) antimatter. The latent building block of $\mathscr{G}$ is an irreducible (co-)representation of $\mathbf{Spin}^+(1,3)$: any such block corresponds to a quadruple of neurons per geometric dimension. Under mirror supersymmetry, the instantaneous geometric dimensionality of neural (co-)representations of a (co-)behavioral state is $8$, in the same direction of (co-)flow.

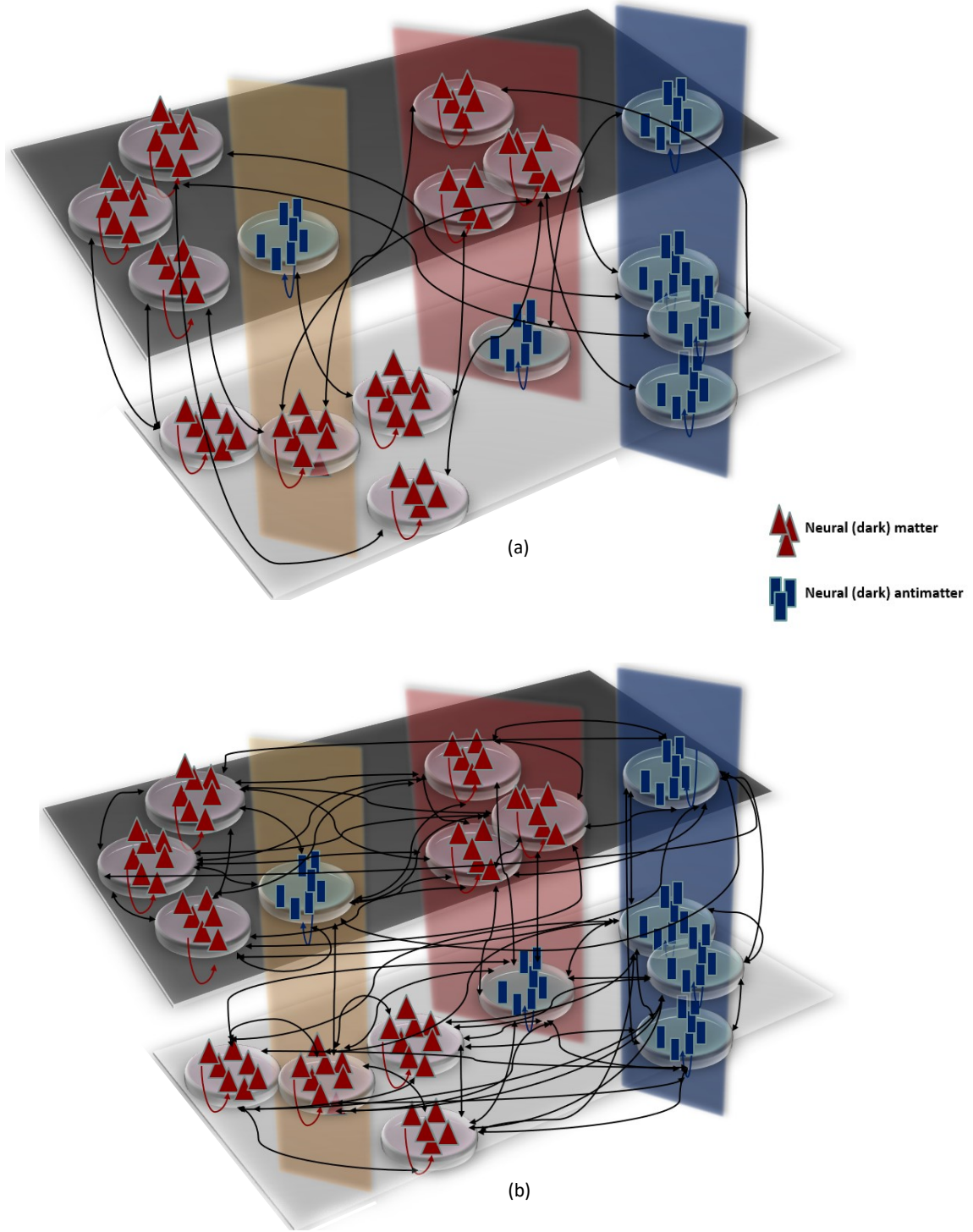

**Fig. 5. Lorentz, (1,4)-crystals.** (a) A Lorentz, $(1,4)$-partition ((b) A Lorentz, co-$(1,4)$-partition) on a bipartition (resp., co-bipartition) of $\mathscr{G}$'s spin-geometrical minor: any two distinct deep clusters contained in same-colored layers are necessarily disconnected (resp., induce a semicomplete subdigraph). Note that $1$ out of $1{+}4$ layers consists exclusively of neural (dark) antimatter. The latent building block of $\mathscr{G}$ is an irreducible (co-)representation of $\mathbf{Spin}^+\left(1,4\right)$: any such block corresponds to an octuple of neurons per geometric dimension. Under mirror supersymmetry, the instantaneous geometric dimensionality of neural (co-)representations of a (co-)behavioral state is $16$, in the same direction of (co-)flow.

**Technical Material (TM) for**

# Relativistic quantum decryption of large-scale neural coding

**Sofia Karamintziou**

**Email:** skaramintziou3@gatech.edu or skaramintziou@iti.gr

# I. Asymmetric Symmetries

Let us introduce a nondegenerate quadratic form, $\vartheta : X \to \mathbb{R}$, on an $\mathbb{R}$-linear space $X$, with $\dim_{\mathbb{R}} X = p + q < \infty$; therewith, $(X, \vartheta)$ is isomorphic to the pseudo-Euclidean space $\mathbb{R}^{p,q}$, with signature $(p, q)$ and positive index $p$.

We denote by $\mathfrak{T}(X) := \bigoplus_{d=0}^{\infty} X^{\otimes d} = \mathbb{R} \oplus X \oplus X^{\otimes 2} \oplus \dots$ the tensor algebra on $X$, and by $\mathrm{I}(\vartheta) := \left\langle x \otimes x - \vartheta(x) \cdot 1_{\mathfrak{T}(X)} : x \in X \right\rangle \subseteq \mathfrak{T}(X)$ the two-sided ideal in $\mathfrak{T}(X)$ generated by all elements of the form $x \otimes x - \vartheta(x) \cdot 1_{\mathfrak{T}(X)}$; $x \in X$. This algebra $\mathfrak{T}(X)$ admits a natural $\mathbb{Z}_2$-grading $\mathfrak{T}(X) = \mathfrak{T}^0(X) \oplus \mathfrak{T}^1(X)$, where $\mathfrak{T}^0(X) := \bigoplus_{d=0}^{\infty} X^{\otimes 2d}$ and $\mathfrak{T}^1(X) := \bigoplus_{d=0}^{\infty} X^{\otimes(2d+1)}$. Recursively, one defines a $\mathbb{Z}_2$-graded algebra structrure on the even tensor algebra, $\mathfrak{T}^0(X)$.

Let $\mathrm{I}^0(\vartheta) := \mathrm{I}(\vartheta) \cap \mathfrak{T}^0(X)$. The quotient algebra $\mathfrak{C}^0(X, \vartheta) := \mathfrak{T}^0(X) / \mathrm{I}^0(\vartheta)$ is the *even Clifford algebra of* $(X, \vartheta)$ and will generally be abbreviated by $\mathfrak{C}^0_{p,q}$.

The *reversal* or *transpose* map $^{\tau} : \mathfrak{T}^0(X) \to \mathfrak{T}^0(X)$ given by $x_1 \otimes x_2 \otimes \dots \otimes x_k \mapsto x_k \otimes \dots \otimes x_2 \otimes x_1$ on pure tensors in $\mathfrak{T}^0(X)$ (and extended $\mathbb{R}$-linearly) preserves $\mathrm{I}^0(\vartheta)$, and thereby descends to an anti-involution on $\mathfrak{C}^0_{p,q}$.

By construction, $\mathfrak{C}^0_{p,q}$ retains a $\mathbb{Z}_2$-grading as a real space, namely $\mathfrak{C}^0_{p,q} = \mathfrak{C}^{0[0]}_{p,q} \oplus \mathfrak{C}^{0[1]}_{p,q}$, where $\mathfrak{C}^{0[0]}_{p,q}$ is the image of $\bigoplus_{d=0}^{\infty} X^{\otimes 2(2d)}$ under the projection $\mathfrak{T}^0(X) \to \mathfrak{C}^0_{p,q}$, while $\mathfrak{C}^{0[1]}_{p,q}$ is the image of $\bigoplus_{d=0}^{\infty} X^{\otimes 2(2d+1)}$ under the same map. This grading on $\mathfrak{C}^0_{p,q}$ may as well be defined in terms of the reversal operation on $\mathfrak{C}^0_{p,q}$: $\mathfrak{C}^{0[i]}_{p,q} := \left\{ x \in \mathfrak{C}^0_{p,q} : x^{\tau} = (-1)^i x \right\}$, $i = 0, 1$.

We now let $\{e_1, \dots, e_p, \dots, e_{p+q}\}$ be an orthonormal basis for $X$, and we identify $X^{\otimes 2}$ with its image under the canonical injection $X^{\otimes 2} \to \mathfrak{T}^0(X)$. Then, as an $\mathbb{R}$-algebra, $\mathfrak{C}^0_{p,q}$ is generated by the set $\{e_{i_1} \cdot e_{i_2} : 1 \leq i_1 < i_2 \leq p + q\}$ and, as an $\mathbb{R}$-linear space, it has a basis $\mathcal{B} := \{e_{i_1} \cdot e_{i_2} \cdots e_{i_r} : 1 \leq i_1 < \dots < i_r \leq p + q; r \text{ even}\}$, the empty product ($r = 0$) being the multiplicative identity element $1_{\mathfrak{C}^0_{p,q}}$. As such, $\dim_{\mathbb{R}} \mathfrak{C}^0_{p,q} = 2^{p+q-1}$. For simplicity of notation, we will occasionally identify $\mathcal{B}$ with the set $\mathcal{B} = \{b_i\}_{\mathcal{J}}$, where $\mathcal{J} := \{1, \dots, 2^{p+q-1}\}$. As well, we shall consider the partition, $\{\mathcal{J}_i\}_{\mathbb{Z}_2}$, of the index set $\mathcal{J}$ associated to the super-space structure of $\mathfrak{C}^0_{p,q}$. In $\dim_{\mathbb{R}} X = 3$, $\mathcal{J}_0$ is a singleton, while $\mathcal{J}_1$ has cardinality 3; in $\dim_{\mathbb{R}} X = 6$, the asymmetry in the partition vanishes non-trivially.

DEFINITION (**quadratic norm**) The *quadratic norm* on the even Clifford algebra $\mathfrak{C}^0_{p,q}$ is the function $\delta : \mathfrak{C}^0_{p,q} \to \mathfrak{C}^0_{p,q}$, given by $x \mapsto x \cdot x^{\tau}$.

The pair $(\mathfrak{C}^0_{p,q}, \delta)$ will be referred to as a *special quadratic algebra* of $(X, \vartheta)$. We will sometimes abbreviate $(\mathfrak{C}^0_{p,q}, \delta)$ merely by $\mathfrak{C}^0_{p,q}$. Whenever $\dim_{\mathbb{R}} X = p + q \leq 3$, the quadratic norm $\delta$ on $\mathfrak{C}^0_{p,q}$ may be identified with a nondegenerate quadratic form, $\tilde{\vartheta} : \mathfrak{C}^0_{p,q} \to \mathbb{R}$, on $\mathfrak{C}^0_{p,q}$. In any dimension, this quadratic form defines $(\mathfrak{C}^0_{p,q}, \tilde{\vartheta})$ as a real nondegenerate quadratic space, isomorphic to the Euclidean space $\mathbb{R}^{2u}$ if $\tilde{\vartheta}$ is anisotropic, and to the pseudo-Euclidean space $\mathbb{R}^{u,u}$ if $\tilde{\vartheta}$ is isotropic, where $u := \dim_{\mathbb{R}} \mathfrak{C}^0_{p,q} / 2$. In $\dim_{\mathbb{R}} X = p + q = 4$, we have $\delta(x) \in \mathcal{Z}(\mathfrak{C}^0_{p,q})$, with $x \in \mathfrak{C}^0_{p,q}$ and $\mathcal{Z}(\mathfrak{C}^0_{p,q})$ denoting the center of $\mathfrak{C}^0_{p,q}$.

Since $\left(\mathfrak{C}^0_{p,q}, \tilde{\vartheta}\right)$ is a nondegenerate real quadratic space, there exists an isometry, $\tilde{\vartheta} \simeq \tilde{\vartheta}^0 \perp \tilde{\vartheta}^1$, from $\tilde{\vartheta}$ to the orthogonal direct sum $\tilde{\vartheta}^0 \perp \tilde{\vartheta}^1$, with $\tilde{\vartheta}^0$ positive-definite and $\tilde{\vartheta}^1$ negative-definite. This isometry uniquely defines another partition of the index set $\mathcal{J}$; we shall write $\left\{\mathcal{J}^\dagger_i\right\}_{\mathbb{Z}_2}$ to denote this partition, so as to distinguish it from the partition $\left\{\mathcal{J}_i\right\}_{\mathbb{Z}_2}$ of $\mathcal{J}$. Of course, $\left(\mathfrak{C}^0_{p,q}, \tilde{\vartheta}\right)$ is isotropic if and only if $\left\{\mathcal{J}^\dagger_i\right\}_{\mathbb{Z}_2}$ is nontrivial; if this is the case, then, by the previous remarks, $\left(\mathfrak{C}^0_{p,q}, \tilde{\vartheta}\right)$ is a hyperbolic space.

For $\dim_\mathbb{R} X = p + q \geq 1$, let $\mathfrak{C}^{0\times}_{p,q}$ denote the multiplicative group of invertible elements in $\mathfrak{C}^0_{p,q}$, and let $\boldsymbol{\Gamma}^0_{p,q} \supset \mathbf{Spin}\left(p,q\right)$ denote the *special Clifford-Lipschitz group* in $\mathfrak{C}^{0\times}_{p,q}$, the subgroup $\mathbf{Spin}\left(p,q\right)$ being normal. In particular, $\boldsymbol{\Gamma}^0_{p,q}$ and $\mathbf{Spin}\left(p,q\right)$ are the subgroups of $\mathfrak{C}^{0\times}_{p,q}$, for which the following short exact sequences, respectively, exist

$$1 \to \mathbb{R}^\times \to \boldsymbol{\Gamma}^0_{p,q} \to \mathrm{SO}\left(p,q\right) \to 1 \tag{1.a}$$

$$1 \to \mathbb{Z}_2 \to \mathbf{Spin}\left(p,q\right) \to \mathrm{SO}\left(p,q\right) \to 1 \tag{1.b}$$

the epimorphisms $\boldsymbol{\Gamma}^0_{p,q} \to \mathrm{SO}\left(p,q\right)$ and $\mathbf{Spin}\left(p,q\right) \to \mathrm{SO}\left(p,q\right)$ being restrictions of the adjoint representation of $\mathfrak{C}^{0\times}_{p,q}$. These epimorphisms induce the short exact sequences

$$1 \to \mathbb{R}^\times \to \boldsymbol{\Gamma}^{0+}_{p,q} \to \mathrm{SO}^+\left(p,q\right) \to 1 \tag{2.a}$$

$$1 \to \mathbb{Z}_2 \to \mathbf{Spin}^+\left(p,q\right) \to \mathrm{SO}^+\left(p,q\right) \to 1 \tag{2.b}$$

for the identity components of $\boldsymbol{\Gamma}^0_{p,q}$ and $\mathbf{Spin}\left(p,q\right)$, respectively.

DEFINITION (**null super-cone**) The vanishing locus, $\mathcal{O}\left(\delta\right) := \left\{x \in \mathfrak{C}^0_{p,q} : \delta\left(x\right) = 0_{\mathfrak{C}^0_{p,q}}\right\}$, of the quadratic norm, $\delta$, on the even Clifford algebra $\mathfrak{C}^0_{p,q}$ is the *null super-cone* of $\delta$.

**Proposition 1.** *In* $\dim_\mathbb{R} X = p + q \leq 5$,

$$\mathcal{O}\left(\delta\right) = \mathcal{O}^0_{p,q}.$$

*That is, the null super-cone,* $\mathcal{O}\left(\delta\right)$*, of the quadratic norm,* $\delta$*, on the even Clifford algebra,* $\mathfrak{C}^0_{p,q}$*, of* $\left(X, \vartheta\right)$ *is self-dual to the common boundary,* $\mathcal{O}^0_{p,q}$*, of the disjoint open sets* $\boldsymbol{\Gamma}^{0+}_{p,q}$ *and* $\boldsymbol{\Gamma}^0_{p,q} \setminus \boldsymbol{\Gamma}^{0+}_{p,q}$*. The statement is not true in* $\dim_\mathbb{R} X = p + q = 6$*, with* $p \neq 0$*, where mirror supersymmetry breaks down.*

*Proof.* In $\dim_\mathbb{R} X = p + q \leq 5$, we have $\boldsymbol{\Gamma}^0_{p,q} = \left\{x \in \mathfrak{C}^0_{p,q} : \delta\left(x\right) \in \mathbb{R}^\times \cdot 1_{\mathfrak{C}^0_{p,q}}\right\}$, while $\boldsymbol{\Gamma}^{0+}_{p,q} = \left\{x \in \mathfrak{C}^0_{p,q} : \delta\left(x\right) \in \mathbb{R}^{\times 2} \cdot 1_{\mathfrak{C}^0_{p,q}}\right\}$, $\mathbb{R}^\times$ and $\mathbb{R}^{\times 2}$ denoting the non-zero and positive real scalars, respectively. In $\dim_\mathbb{R} X = p + q = 6$, $\boldsymbol{\Gamma}^{0+}_{p,q} \subset \left\{x \in \mathfrak{C}^0_{p,q} : \delta\left(x\right) \in \mathbb{R}^{\times 2} \cdot 1_{\mathfrak{C}^0_{p,q}}\right\}$, i.e., $\boldsymbol{\Gamma}^{0+}_{p,q}$ is a proper subset of $\left\{x \in \mathfrak{C}^0_{p,q} : \delta\left(x\right) \in \mathbb{R}^{\times 2} \cdot 1_{\mathfrak{C}^0_{p,q}}\right\}$. ∎

The form of self-duality observed in proposition I.1, reminiscent of the illusion elaborated in [T.1], we refer to as *Wittgenstein-* or $\mathcal{W}$-*duality* under mirror super(a)symmetry. One may speculate that there is a link to the theory of *spinning black-hole binaries* [T.2].

The epimorphism $\mathbf{Spin}^+(p,q) \rightarrow \mathrm{SO}^+(p,q)$ in (2.b) defines the standard $(p+q)$-dimensional representation of the identity component of the spin group on X. On the other hand, any (left) irreducible real representation, $\mathfrak{C}^0_{p,q} \rightarrow \mathrm{End}_{\mathbb{R}}(W)$, of the even Clifford algebra $\mathfrak{C}^0_{p,q}$ restricts to an irreducible representation, $\mathbf{Spin}^+(p,q) \rightarrow \mathrm{Aut}_{\mathbb{R}}(W)$, of $\mathbf{Spin}^+(p,q)$, referred to as a *spinor representation* on the $\mathbb{R}$-module W. This module W admits a natural super-space structure, $W := W^0 \oplus W^1$, induced by the respective structure of $\mathfrak{C}^0_{p,q} = \mathfrak{C}^{0[0]}_{p,q} \oplus \mathfrak{C}^{0[1]}_{p,q}$.

In general, the set, $\mathrm{Rep}\left(\mathfrak{C}^0_{p,q}\right)$, of real representations of the even Clifford algebra $\mathfrak{C}^0_{p,q}$ forms an abelian monoid under the direct sum, $\oplus$. The group completion of this monoid is the Grothendieck group, $\mathrm{K}\left(\mathfrak{C}^0_{p,q}\right)$, of $\mathfrak{C}^0_{p,q}$ — the free abelian group generated by the isomorphism classes of irreducible real representations of $\mathfrak{C}^0_{p,q}$ (hence, by the isomorphism classes of real spinor representations) — together with the monoid map $[\ \ ] : \mathrm{Rep}\left(\mathfrak{C}^0_{p,q}\right) \rightarrow \mathrm{K}\left(\mathfrak{C}^0_{p,q}\right)$. Specifically, if $\mathfrak{C}^0_{p,q}$ is a simple algebra, we obtain the isomorphism $\mathrm{K}\left(\mathfrak{C}^0_{p,q}\right) \simeq \mathbb{Z}$ of abelian groups, sending $[W]$ to 1 (the length of W as a finitely generated projective $\mathfrak{C}^0_{p,q}$-module); otherwise, $\mathrm{K}\left(\mathfrak{C}^0_{p,q}\right) \simeq \mathbb{Z} \oplus \mathbb{Z}$. Under the tensor product, $\otimes_{\mathbb{R}}$, $\mathrm{Rep}\left(\mathfrak{C}^0_{p,q}\right)$ possesses the structure of a commutative semiring; therewith, $\mathrm{K}\left(\mathfrak{C}^0_{p,q}\right)$ is turned to a commutative ring — the *Grothendieck ring* of $\mathfrak{C}^0_{p,q}$.

We refer the interested reader to [T.3-15] for additional bibliography related to this section.

## II. The Flow Network

**II.1** We start by considering a finite set of large cardinality, $\mathcal{I} := \left\{1,2,\ldots,n : n \text{ large}\right\}$, indexing agents whose interactions over a *flow network* $\mathcal{F}$ are represented by an ordered triple, $\mathcal{F} := \left(\mathcal{D}_\gamma, \mathcal{R}, \mathcal{F}\right)$, consisting of:

1. a weighted *spin geometry*, namely a non-simple finite digraph, $\mathcal{D}_\gamma := \left(\mathcal{D}_\gamma \, ; \, \sigma \circ \kappa, \, \tau \circ \kappa\right)$, where $\mathcal{D}_\gamma := (\mathcal{D}, \gamma)$ denotes the digraph with *trivial* structure associated to $\mathcal{D}_\gamma$.

   In particular, in the digraph $\mathcal{D} := (\mathcal{V}, \mathcal{E})$, we let $\mathcal{V} := \left\{v_i\right\}_{\mathcal{I}}$ be the set of its $n$ *vertices*, the $i$-th element of which represents the $i$-th agent in the network, and $\mathcal{E} := \left\{e_e\right\}_{\tilde{\mathcal{I}}} \subseteq \mathcal{V} \times \mathcal{V}$, with $\tilde{\mathcal{I}} := \left\{1,2,\ldots,m\right\}$, be a reflexive relation on $\mathcal{V}$. The $e$-th ordered pair of vertices, $e_e = \left(v_i, v_j\right)$, in $\mathcal{E}$ represents an *arc*, or *edge* directed from source vertex $v_i$ to sink vertex $v_j$ such that $v_i$ has an *out-neighbor* $v_j$, and $v_j$ has an *in-neighbor* $v_i$. For $i = j$, $e_e$ represents an edge such that $v_i \in \mathcal{V}$ is both an out-neighbor and in-neighbor of itself, i.e., a *self-loop*. We shall denote by $\mathcal{N}_i^{in} := \left\{j \in \mathcal{I} : \left(v_j, v_i\right) \in \mathcal{E}\right\}$ and $\mathcal{N}_i^{out} := \left\{j \in \mathcal{I} : \left(v_i, v_j\right) \in \mathcal{E}\right\}$ the sets of indices of, accordingly, the in- and out-neighbors of vertex $v_i$ in $\mathcal{D}$. Each $e_e$ in $\mathcal{E}$, other than a self-loop, represents a link between two agents due to directed influence. Self-loops in $\mathcal{D}$ reflect the intrinsic properties of an agent. The map $\gamma : \mathcal{V} \times \mathcal{V} \rightarrow \mathbb{R}_0$ in $\mathcal{D}_\gamma$ assigns to every $\left(v_i, v_j\right) \in \mathcal{V} \times \mathcal{V}$ a unique real value denoted by $\gamma_{ij}$; this value is strictly positive if $\left(v_i, v_j\right) \in \mathcal{E}$, and 0 if $\left(v_i, v_j\right) \notin \mathcal{E}$. For $i \neq j$, $\gamma_{ij}$ quantifies the strength of the associated influence. Inherent in $\mathcal{D}_\gamma$ are several partitions. First, the partition, $\left\{\mathcal{I}_c\right\}_{\mathcal{C}}$, of $\mathcal{I}$, with $\mathcal{C} := \left\{1,2,\ldots,k\right\}$, associated to the partition, $\tilde{\mathcal{V}} := \left\{\mathcal{V}_c\right\}_{\mathcal{C}}$, of $\mathcal{V}$ into $k \geq 2$ subsets of large cardinality referred to as *deep clusters*; elements of $\mathcal{V}_c := \left\{v_i\right\}_{\mathcal{I}_c}$, with $c \in \mathcal{C}$, are referred to as $c$-*vertices*. Therewith, a unique equivalence relation $\sim$ is defined on $\mathcal{V}$, whose set, $\mathcal{V}/\sim$, of equivalence classes is exactly $\tilde{\mathcal{V}}$. We let $\kappa : \mathcal{V} \to \mathcal{V}/\sim$ be the natural projection sending each element of $\mathcal{V}$ to its equivalence class. The map $\sigma : \tilde{\mathcal{V}} \to \mathbb{Z}_2$ assigns to every $\mathcal{V}_i$ in $\tilde{\mathcal{V}}$ a unique value $\sigma_i$ in $\mathbb{Z}_2$, thereby inducing a partition of the index set $\mathcal{C}$ itself as $\left\{ \mathcal{C}_i \right\}_{\mathbb{Z}_2}$ and of the set, $\tilde{\mathcal{V}}$, of deep clusters as $\left\{ \tilde{\mathcal{V}}_i \right\}_{\mathbb{Z}_2}$. We think of the value, $\sigma \circ \kappa \left( v_i \right)$, of the composition of maps $\kappa$ and $\sigma$ at $v_i$ as indicating whether the influence exerted by an agent represented by $v_i$ in $\mathcal{V}_c$ is, in a sense, *proactive* or *counteractive*. On the other hand, the map $\tau : \tilde{\mathcal{V}} \to \mathbb{Z}_2$ gives rise to the partitions $\left\{ \mathcal{C}_i^\dagger \right\}_{\mathbb{Z}_2}$ of $\mathcal{C}$ and $\left\{ \tilde{\mathcal{V}}_i^\dagger \right\}_{\mathbb{Z}_2}$ of $\tilde{\mathcal{V}}$.

Let us denote by $\boldsymbol{\mathcal{D}} := \left( \mathcal{D} \, ; \, \sigma \circ \kappa, \, \tau \circ \kappa \right)$ the unweighted geometry in $\boldsymbol{\mathcal{D}}_i$.

DEFINITION **(geometrical minor)** Let $\tilde{\mathcal{D}} := \left( \tilde{\mathcal{V}}, \tilde{\mathcal{E}} \right)$ be the image of $\mathcal{D}$ under the epimorphism $\kappa : \mathcal{V} \to \mathcal{V}/\sim$ of digraphs. The digraph $\tilde{\boldsymbol{\mathcal{D}}} := \left( \tilde{\mathcal{D}} \, ; \, \sigma, \tau \right)$ is the *geometrical minor* of $\boldsymbol{\mathcal{D}}$.

REMARK 1: A $j$-*coloring* of $\tilde{\mathcal{D}}$ is a surjective map $\mu : \tilde{\mathcal{V}} \to \{1, 2, \ldots, j\}$ such that each subset $\mu^{-1}(i)$ of vertices is *independent* (that is, no two distinct vertices in the subset belong to the same ordered pair in $\tilde{\mathcal{E}}$). Clearly, if $\tilde{\mathcal{D}}$ is a tournament, the minimum $j$ for which $\tilde{\mathcal{D}}$ has a $j$-coloring is $k$, and $\tilde{\mathcal{D}}$ is $k$-*chromatic*.

$\mathcal{D}$ is *bipartite* if there exists a partition of $\mathcal{V}$ into two nonempty subsets $\mathcal{V}^i$ and $\mathcal{V}^{i+1}$, such that every edge of $\mathcal{D}$ is either a self-loop or an edge directed from a source vertex in $\mathcal{V}^i$ to a sink vertex in $\mathcal{V}^{i+1}$; $i \in \mathbb{Z}_2$. If this is the case, we say that $\mathcal{D}$ has *bipartition* $\left( \mathcal{V}^i, \mathcal{V}^{i+1} \right)$. A *co-bipartition*, $\left( \mathcal{V}^i, \mathcal{V}^{i+1} \right)^\perp$, of $\mathcal{D}$ is the structurally dual (complementary) construction [T.16, T.17].

DEFINITION **(Lorentz partition)** A $\left( 1, q \right)$-*partition* of $\mathcal{D}$ is a partition of $\mathcal{V}$ into $1 + q$ nonempty subsets $\mathcal{V}^0, \mathcal{V}^1, \ldots, \mathcal{V}^q$ such that

*(i)* $\mathcal{V}^0 = \cup \mathcal{V}_i^0$, with $\mathcal{V}_i^0 \in \tilde{\mathcal{V}}_0$, and,

*(ii)* every edge of $\mathcal{D}$ is either a self-loop or an edge directed from a source vertex in $\mathcal{V}^i$ to a sink vertex in $\mathcal{V}^j$; $i \neq j$.

We say that $\mathcal{D}$ is $\left( 1, q \right)$-*partite* or that $\mathcal{D}$ has *Lorentz partition* $\left( \mathcal{V}^0, \cdot \right)$. By structural duality, one appropriately defines a *Lorentz co-partition* or *co-*$\left( 1, q \right)$-*partition* of $\mathcal{D}$.

A *subdigraph*, $\hat{\mathcal{D}} := \left( \hat{\mathcal{V}}, \hat{\mathcal{E}} \right)$, of $\mathcal{D}$ will be a digraph with set of vertices $\hat{\mathcal{V}} \subseteq \mathcal{V}$ and set of edges $\hat{\mathcal{E}} \subseteq \hat{\mathcal{V}} \times \hat{\mathcal{V}} \subseteq \mathcal{E}$. If $\hat{\mathcal{V}} = \mathcal{V}$, we say that $\hat{\mathcal{D}}$ is a *spanning subdigraph* of $\mathcal{D}$, which we occasionally denote by $\mathcal{D} | \hat{\mathcal{E}}$; in the case where $\hat{\mathcal{E}}$ is nonempty, $\mathcal{D} | \hat{\mathcal{E}}$ is a *factor* of $\mathcal{D}$. We say that $\hat{\mathcal{D}}$ is *induced by* $\hat{\mathcal{V}}$ or that $\hat{\mathcal{D}}$ is an *induced subdigraph*, $\mathcal{D} \langle \hat{\mathcal{V}} \rangle$, if it is the maximal subdigraph of $\mathcal{D}$ with set of vertices $\hat{\mathcal{V}}$.

We let $\mathcal{E}_c := \left\{ \left( v_i, v_j \right) \in \mathcal{E} : v_i \in \mathcal{V}_c \right\}$ be the set of all edges in $\mathcal{D}$ whose source vertex is a $c$-vertex, hereafter called the *$c$-relation on* $\mathcal{V}$ or the set of $c$-*edges*. For each $c \in \mathcal{C}$ and $i \in \mathcal{I}$, $\mathcal{N}_i^{in(c)} := \left\{ j \in \mathcal{N}_i^{in} : v_j \in \mathcal{V}_c \right\}$ denotes the set of indices of those vertices in $\mathcal{D}$, from which there is a $c$-edge to $v_i$. Likewise, $\mathcal{N}_i^{out(c)} := \left\{ j \in \mathcal{N}_i^{out} : v_i \in \mathcal{V}_c \right\}$; so if $v_i \in \mathcal{V}_t$, $\mathcal{N}_i^{out(c)} = \mathcal{N}_i^{out}$ for $c = t$, and $\mathcal{N}_i^{out(c)} = \emptyset$ for all $c \neq t$.

A $c-factor$, $\mathcal{D}\big|_{\mathcal{E}_c}$, of $\mathcal{D}$ is a factor of $\mathcal{D}$ with set of edges $\mathcal{E}_c$. As a natural consequence

**Proposition 1.** *Under the geometry* $\mathcal{D}$ , $\mathcal{D}$ *is the direct sum of the indexed family* $\left(\mathcal{D}\big|_{\mathcal{E}_c}\right)_{\mathcal{C}}$ *of* $c-factors$. *We write*

$$\mathcal{D} \simeq \bigoplus_{\mathcal{C}} \mathcal{D}\big|_{\mathcal{E}_c} \simeq \bigoplus_{i=0,1}\left(\bigoplus_{\mathcal{C}_i} \mathcal{D}\big|_{\mathcal{E}_c}\right).$$

REMARK 2: One obtains the directional dual to proposition 1, by defining $\mathcal{E}_c^{\uparrow}$ as the set of all edges in $\mathcal{D}$ whose *sink* vertex is a $c$-vertex.

For us, the *vertex module*, $\mathscr{V}$ , of $\mathcal{D}$ will be the free $\mathfrak{C}_{p,q}^0$ - module $\mathscr{V} := \mathfrak{C}_{p,q}^{0~(\mathcal{V})} \simeq \bigoplus_{v\in\mathcal{V}} \mathfrak{C}_{p,q}^0 v$ , where we identify $\mathcal{V}$ with its image under the canonical injection $\mathcal{V} \to \mathscr{V}$ . Since $\mathfrak{C}_{p,q}^0$ is a Noetherian ring, we may write $\text{rank}_{\mathfrak{C}_{p,q}^0} \mathscr{V} = n$ . In a similar way, we define the *edge module*, $\mathfrak{E}$ , of $\mathcal{D}$ by $\mathfrak{E} := \mathfrak{C}_{p,q}^{0~(\mathcal{E})} \simeq \bigoplus_{e\in\mathcal{E}} \mathfrak{C}_{p,q}^0 e$ , which is obtained as a free submodule of $\mathscr{V}^{\otimes 2}$ . We refer to the ordered pair $\left(\mathscr{V}, \mathfrak{E}\right)$ as the *module pair* of $\mathcal{D}$ . The module pair $\left(V, E\right)$ of $\mathcal{D}$ is the pair of $\mathbb{R}$ - modules for which the isomorphisms $\mathscr{V} \simeq \mathfrak{C}_{p,q}^0 \otimes_{\mathbb{R}} V$ and $\mathfrak{E} \simeq \mathfrak{C}_{p,q}^0 \otimes_{\mathbb{R}} E$ , as free $\mathfrak{C}_{p,q}^0$ - modules, exist.

For the *$c$-vertex module*, $V_c$ , of $\mathcal{D}$ we write $V_c := \mathbb{R}^{(\mathcal{V}_c)} \simeq \bigoplus_{v\in\mathcal{V}_c}\mathbb{R}\,v$ .

DEFINITION (**$c$ − edge module**) Let $V_c \otimes_{\mathbb{R}} V \subseteq V^{\otimes 2}$ be the tensor product of $\mathbb{R}$-modules $V_c$ and $V$ , i.e., the $\mathbb{R}$-module with basis $\left\{v_i \otimes v_j\right\}_{(i,j)\in\mathcal{I}_c\times\mathcal{I}}$ . The *$c$-edge module*, $E_c$ , of $\mathcal{D}$ is obtained as a (free) submodule of $V_c \otimes_{\mathbb{R}} V$ . In particular, $E_c := \mathbb{R}^{(\mathcal{E}_c)} \simeq \bigoplus_{e\in\mathcal{E}_c}\mathbb{R}\,e$ .

By the previous remarks, under the geometry $\mathcal{D}$ , $\left(V,E\right) \simeq \left(\bigoplus_{\mathcal{C}} V_c, \bigoplus_{\mathcal{C}} E_c\right) \simeq \left(\bigoplus_{i=0,1}\left(\bigoplus_{\mathcal{C}_i} V_c\right), \bigoplus_{i=0,1}\left(\bigoplus_{\mathcal{C}_i} E_c\right)\right)$. Further

**Proposition 2.** *The epimorphism of digraphs* $\kappa : \mathcal{D} \to \tilde{\mathcal{D}}$ *induces an* $\mathbb{R}$-*epimorphism* $\left(V,E\right) \to \left(\tilde{V},\tilde{E}\right)$, *where* $\left(\tilde{V},\tilde{E}\right)$ *denotes the module pair of* $\tilde{\mathcal{D}}$.

We require that the $\mathbb{R}$-module $\tilde{V}$ possess an $\mathbb{R}$-algebra structure.

**Axiom 1.** *(i)* *There is a real nondegenerate quadratic space* $\left(X, \vartheta\right)$ *and a basis for* $X$ , *such that* $\tilde{V}$ *is identified with the even Clifford algebra,* $\mathfrak{C}_{p,q}^0$ , *of* $\left(X,\vartheta\right)$. *In particular,* $\mathcal{C} = \mathcal{I}$ , $\mathcal{C}_i = \mathcal{I}_i$ , *and* $\mathcal{C}_i^{\uparrow} = \mathcal{I}_i^{\uparrow}$ ; $i \in \mathbb{Z}_2$ .

*(ii)* *The module pair* $\left(\mathscr{V}, \mathfrak{E}\right)$ *of* $\mathcal{D}$ *is identified with the pair of free* $\tilde{V}$-*modules* $\left(\tilde{V}\otimes_{\mathbb{R}} V, \tilde{V}\otimes_{\mathbb{R}} E\right)$, *so* $\left(\mathscr{V}, \mathfrak{E}\right)$ *corresponds to a pair of elements in the Grothendieck ring,* $\mathrm{K}\left(\tilde{V}\right)$ , *of* $\tilde{V}$.

*(iii)* *There is an epimorphism* $\mathcal{D} \to \mathcal{D}$ *of digraphs inducing an epimorphism* $\left(\mathscr{V}, \mathfrak{E}\right) \to \left(V,E\right)$ *of projective (left)* $\tilde{V}$-*module pairs; therewith,* $\left(V,E\right)$ *corresponds to a pair of elements in the Grothendieck ring,* $\mathrm{K}\left(\tilde{V}\right)$ , *of* $\tilde{V}$.

REMARK 3: By Axiom 1(*iii*), the (pairs of) direct summands in the decomposition $\left(V,E\right) \simeq \left(\bigoplus_{\mathcal{C}} V_c, \bigoplus_{\mathcal{C}} E_c\right)$ are turned to (pairs of) projective (left) $\tilde{V}$-modules.

The notations $\delta$ and $\tilde{\vartheta}$ will once again be used to denote, accordingly, the quadratic norm and quadratic form on $\tilde{V}$. The notation $\tilde{V}_{\mathcal{F}}$ in place of $\tilde{V}$ will occasionally be useful. We usually write $\left({}_V V, {}_V E\right)$ when referring to $\left(V, E\right)$ as a left $\tilde{V}$-module pair.

We define $\underline{\mathcal{E}}$ to be the set of edges of *the underlying graph*, $\mathcal{G} := \left(\mathcal{V}, \underline{\mathcal{E}}\right)$, of $\mathcal{D}$; that is, $\underline{\mathcal{E}}$ is the set of all elements under the reflexive and symmetric relation generated by $\mathcal{E}$. An edge, $e_{ij} = \left\{v_i, v_j\right\}$, in $\underline{\mathcal{E}}$ is to be thought of as consisting of two *half edges*, denoted by $\bar{e}_{ij}$ and $\bar{e}_{ij}$, and the set of all half-edges in $\mathcal{G}$ is denoted by $\underline{\mathcal{E}}$.

Consider the multiplicative group $B := \left\{\pm b_j\right\}_{j \in \mathcal{J}}$ in $\mathfrak{C}_{p,q}^\vartheta \simeq \tilde{V}$.

**Proposition 3.** *There is a unique map* $\omega : \mathcal{V} \times \underline{\mathcal{E}} \to B \cup \{0\}$ *such that* $\mathcal{D} = \left(\mathcal{G}, \omega\right)$.

*Proof.* For $a \in \mathcal{I}$, we define $\omega$ to be the map assigning to every ordered pair $\left(v_a, \bar{e}_{ij}\right)$ $\left(\left(v_a, \tilde{e}_{ij}\right)\right)$ in $\mathcal{V} \times \underline{\mathcal{E}}$, a unique element in $B \cup \{0\}$, denoted by $\omega\left(v_a, \bar{e}_{ij}\right)$ (resp., $\omega\left(v_i, \tilde{e}_{ij}\right)$), subject to the relations $\omega\left(v_i, \bar{e}_{ij}\right) + \omega\left(v_i, \bar{e}_{ii}\right) = 0$; $\omega\left(v_i, \bar{e}_{ij}\right) + \omega\left(v_j, \bar{e}_{ij}\right) = 0$ (resp., $\omega\left(v_i, \tilde{e}_{ij}\right) + \omega\left(v_j, \tilde{e}_{ij}\right) = 0$), $i \ne j$. In particular, if $\omega\left(v_i, \bar{e}_{ij}\right) = b_c$ $\left(\omega\left(v_j, \tilde{e}_{ij}\right) = b_c\right)$ with $c \in \mathcal{J}$, then $\left(v_i, v_j\right) \in \mathcal{E}$ with $v_i \in \mathcal{V}_c$ (resp., $\left(v_j, v_i\right) \in \mathcal{E}$ with $v_j \in \mathcal{V}_c$), while if $\omega\left(v_i, \bar{e}_{ij}\right) = 0$ $\left(\omega\left(v_j, \tilde{e}_{ij}\right) = 0\right)$, $\left(v_i, v_j\right) \notin \mathcal{E}$ (resp., $\left(v_j, v_i\right) \notin \mathcal{E}$). And for all $a \ne i, j$, $\omega\left(v_a, \tilde{e}_{ij}\right) = \omega\left(v_a, \bar{e}_{ij}\right) = 0$. This mapping $\omega$ is unique up to permutation of $\mathcal{J}$. ∎

We think of $\omega$ as a *special Clifford biorientation* of $\mathcal{G}$; we sometimes write $\omega_{i \to j}$ $\left(\omega_{j \to i}\right)$ for $\omega\left(v_i, \bar{e}_{ij}\right)$ (resp, $\omega\left(v_j, \tilde{e}_{ij}\right)$), and $\omega_{j \to i}$ $\left(\omega_{i \to j}\right)$ for $\omega\left(v_j, \bar{e}_{ij}\right)$ (resp, $\omega\left(v_i, \tilde{e}_{ij}\right)$).

The homomorphism of free $\tilde{V}$-modules, $\varpi : \mathfrak{C} \to \mathcal{V}\mathcal{V}$, defined by $e_e \mapsto \sum_{a=1}^n \omega\left(v_a, \bar{e}_{ij}\right) v_a$ for every $e_e = \left(v_i, v_j\right) \in \mathcal{E}$ with $i \ne j$, and by $e_e \mapsto 0$ for every $e_e = \left(v_i, v_i\right) \in \mathcal{E}$ is the *incidence map* of $\mathcal{D}$.

*Connectivity*:

A *semiwalk of length* $l \in \mathbb{N}_0$ in $\mathcal{D}$ will be a finite alternating sequence of the form $w := \left(v^0, e^0, v^1, \ldots, v^i, e^i, v^{i+1}, \ldots, v^{l-1}, e^{l-1}, v^l\right)$, such that $v^i \in \mathcal{V}$ and either $e^i = \left(v^i, v^{i+1}\right)$ or $e^i = \left(v^{i+1}, v^i\right)$ (or both, in the case where $e^i$ represents a self-loop). The respective subsets of edges in $w$ we denote by $\vec{\mathcal{E}}\left(w\right) := \left\{\left(v^i, v^{i+1}\right) \in w\right\}$ and $\tilde{\mathcal{E}}\left(w\right) := \left\{\left(v^{i+1}, v^i\right) \in w\right\}$. This semiwalk $w$ is closed whenever $v^0 = v^j$. It is a (directed) *walk* in $\mathcal{D}$ if $\tilde{\mathcal{E}}\left(w\right)$ is either empty or a set of self-loops; a *spanning walk* is a walk with set of vertices $\mathcal{V}\left(w\right) = \mathcal{V}$. A *(semi)path* in $\mathcal{D}$ is a (semi)walk of length $l \in \mathbb{N}_0$, in which all vertices are distinct. A *(semi)cycle* in $\mathcal{D}$ is a closed (semi)walk of length $l \in \mathbb{N}_+$, in which vertices $v^i$, with $0 < i \le l$, are distinct.

Let us say that a vertex $v_j$ is *(semi)reachable* from some vertex $v_i$ in $\mathcal{D}$ if there exists a (semi)path in $\mathcal{D}$ from $v_i$ to $v_j$. A nonempty subset of vertices $\hat{\mathcal{V}} \subseteq \mathcal{V}$ is *non-reachable* from some subset $\mathcal{V} \subseteq \mathcal{V}$ in $\mathcal{D}$ if no vertex in $\hat{\mathcal{V}}$ is reachable from some vertex in $\mathcal{V}$. If $\mathcal{V}$ is empty, $\hat{\mathcal{V}}$ is trivially non-reachable from $\mathcal{V}$.

$\mathcal{D}$ is *weakly connected or weak* (*strongly connected or strong*) whenever any two vertices in $\mathcal{D}$ are mutually semireachable (resp., mutually reachable). $\mathcal{D}$ is *disconnected* if it is not weak. By a *weak (strong) component* of $\mathcal{D}$ we mean a maximal weak (resp. strong) subdigraph of $\mathcal{D}$. Of course, a component, $\mathcal{D}\langle\hat{\mathcal{V}}\rangle$, of $\mathcal{D}$ is strong if and only if $\mathcal{D}\langle\hat{\mathcal{V}}\rangle$ has a closed spanning walk. Then, if a vertex $v_j$ in $\hat{\mathcal{V}}$ is reachable from some vertex $v_i$ in $\mathcal{D}$, we say that $\hat{\mathcal{V}}$ *is reachable from* $v_i$ in $\mathcal{D}$; conversely, if a vertex $v_j$ in $\mathcal{D}$ is reachable from some vertex $v_i$ in $\hat{\mathcal{V}}$, we say that $v_j$ *is reachable from* $\hat{\mathcal{V}}$.

Given $\mathcal{D}$, a partition of $\mathcal{V}$ into two nonempty subsets $\hat{\mathcal{V}}$ and $\mathcal{V}\setminus\hat{\mathcal{V}}$ is a *cut*. Let $\delta_{out}\hat{\mathcal{V}} := \left\{(v_i,v_j)\in\mathcal{E} : v_i\in\hat{\mathcal{V}}, v_j\in\mathcal{V}\setminus\hat{\mathcal{V}}\right\}$ denote the subset of edges in $\mathcal{E}$ with source vertex in $\hat{\mathcal{V}}$ and sink vertex in $\mathcal{V}\setminus\hat{\mathcal{V}}$. The set $\partial_{out}\hat{\mathcal{V}} := \left\{v_j\in\mathcal{V}\setminus\hat{\mathcal{V}} : (v_i,v_j)\in\delta_{out}\hat{\mathcal{V}}\right\}$ is the *out-boundary* of $\hat{\mathcal{V}}$ in $\mathcal{D}$; we define the *in-boundary*, $\partial_{in}\hat{\mathcal{V}}$, of $\hat{\mathcal{V}}$ in $\mathcal{D}$ in an analogous way.

Setting the stage for the *k-person game on* $\mathcal{F}$ (to be introduced in section IV), we require

**Axiom 2.** *For every* $c\in\mathcal{C}$, *the in-boundary,* $\partial_{in}\mathcal{V}_c$, *and out-boundary,* $\partial_{out}\mathcal{V}_c$, *of* $\mathcal{V}_c$ *in* $\mathcal{D}$ *are nonempty.*

Mutual reachability is a canonical equivalence relation $\sim$ on $\mathcal{V}$, which contains the equivalence relation generated by $\mathcal{E}$. We let $\varsigma:\mathcal{V}\to\mathcal{V}/\sim$ be the natural projection sending each element $v_i$ of $\mathcal{V}$ to its equivalence class, $[v_i] := \left\{v_j\in\mathcal{V} : v_j\sim v_i\right\}$, under this relation. Such equivalence class we also refer to as a *strong cluster* in $\mathcal{D}$, and define a *co-strong cluster* in $\mathcal{D}$ by structural duality.

DEFINITION (**strong minor**) The *strong minor* of $\mathcal{D}$ is the image of $\mathcal{D}$ under the epimorphism $\varsigma:\mathcal{V}\to\mathcal{V}/\sim$ of digraphs.

A vertex $[v_i]$ of the strong minor of $\mathcal{D}$ is *initial* (*terminal*) if the set of its in-neighbors (resp. out-neighbors) is the singleton $\left\{[v_i]\right\}$. Since the strong minor of $\mathcal{D}$ has no cycle of length $l>1$, there is at least one initial, as well as at least one terminal vertex in this digraph.

In a weak component, $\hat{\mathcal{D}}$, of $\mathcal{D}$, an (*oriented) spanning tree*, $\mathcal{T}$, of $\hat{\mathcal{D}}$ is a maximal subdigraph of $\hat{\mathcal{D}}$, with no semicycles; a *chord* of $\mathcal{T}$ is an edge in $\hat{\mathcal{D}}$, which is not in $\mathcal{T}$. And the unique semicycle of which exactly one edge, $e_e$, is a chord of $\mathcal{T}$, is said to be the *fundamental semicycle* belonging to $e_e$ (with respect to $\mathcal{T}$). A *(maximal) spanning forest* of $\mathcal{D}$ is a maximal subdigraph of $\mathcal{D}$, with no semicycles; each weak component of this subdigraph is a spanning tree.

**Proposition 4.** *The flow module,* $\mathfrak{S}$, *of* $\mathcal{D}$ *is a free submodule of* $\mathfrak{C}$. *In particular,* $\mathfrak{S}\subseteq\text{Ke}\,\varpi$.

*Proof.* A semicycle $s$ of length $l\in\mathbb{N}_+$ in $\mathcal{D}$ (hence, also in $\mathcal{D}$) can be identified with an element $s = \sum_{e=0}^{l-1} c_e e_e$ in $\mathfrak{C}$, where for $l=1$, $c_0=1$; for $l>1$, $c_e$ is defined as follows:

For $e_e = (v_i,v_j)\in\mathcal{E}$, $c_e := \tilde{\vartheta}\left(\omega_{i\to j}\right)\omega_{i\to j}$ or $c_e := -\tilde{\vartheta}\left(\omega_{i\to j}\right)\omega_{i\to j}$ whenever $e_e\in\vec{\mathcal{E}}(s)$ or $e_e\in\overleftarrow{\mathcal{E}}(s)$, respectively.

The set $\mathcal{S}'$ of all such elements is a generating set of $\mathfrak{S}\subseteq\mathfrak{C}$, and it can be readily verified that $\varpi(s)=0$, for all $s\in\mathcal{S}'$. Hence, $\mathfrak{S}\subseteq\text{Ke}\,\varpi$. The subset, $\mathcal{S}\subseteq\mathcal{S}'$, of fundamental semicycles (with respect to an oriented spanning forest of $\mathcal{D}$) is a basis for $\mathfrak{S}$. Thereby, $\mathfrak{S}\simeq\mathfrak{C}_{p,q}^{0\ (\mathcal{S})}$. ∎

DEFINITION (**repellor/attractor**) A *repellor*, $\mathcal{T}_{\leftarrow s}$, (*attractor*, $\mathcal{T}_{\rightarrow b}$), of $\mathcal{D}$ ( $\mathcal{D}$ ) will be the nonempty spanning subdigraph of $\mathcal{D}$ with no semicycles, where each vertex is reachable from the *source*, $[v_s]$ (resp., where the *basin*, $[v_b]$, is reachable from each vertex). If $\mathcal{D}$ has a repellor, $\mathcal{T}_{\leftarrow s}$ (an attractor, $\mathcal{T}_{\rightarrow b}$), we say that $v_i \in \mathcal{V}$ is *submissive to* the source, $[v_s]$ (resp., to the basin, $[v_b]$) in $\mathcal{D}$ if $v_i \notin [v_s]$ (resp., $v_i \notin [v_b]$).

By structural duality, one defines a *co-repellor*, $\mathcal{T}_{\leftarrow s}^{\perp}$, of $\mathcal{D}$, with *co-source* $[v_s]^{\perp}$; and a *co-attractor*, $\mathcal{T}_{\rightarrow b}^{\perp}$, of $\mathcal{D}$, with *co-basin* $[v_b]^{\perp}$.

Assuming that $\mathcal{D}$ has a repellor, $\mathcal{T}_{\leftarrow s}$, with source $[v_s] \subseteq \mathcal{V}_1 \cup \ldots \cup \mathcal{V}_k$, we write $\boldsymbol{\mathcal{V}}$ for $[v_s]$ and $\mathcal{D}_{\boldsymbol{\mathcal{V}}}$ for $\mathcal{D}\langle \boldsymbol{\mathcal{V}}\rangle$. We then let $\tilde{\mathcal{D}}_{\boldsymbol{\mathcal{V}}} := (\tilde{\mathcal{V}}, \tilde{\mathcal{E}})$ be the image of $\mathcal{D}_{\boldsymbol{\mathcal{V}}}$ under the restriction $\kappa|_{\boldsymbol{\mathcal{V}}}$ of the epimorphism $\kappa : \mathcal{V} \to \mathcal{V}/\sim$ of digraphs to $\boldsymbol{\mathcal{V}}$. This restriction is composable with the mappings $\boldsymbol{\sigma} : \tilde{\mathcal{V}} \to \mathbb{Z}_2$ and $\boldsymbol{\tau} : \tilde{\mathcal{V}} \to \mathbb{Z}_2$ defined analogously to $\sigma$ and $\tau$, respectively.

DEFINITION (**geometrical source-minor**) The digraph $\hat{\mathcal{D}} := (\tilde{\mathcal{D}}_{\boldsymbol{\mathcal{V}}}; \boldsymbol{\sigma}, \boldsymbol{\tau})$ is the *geometrical source-minor* of $\mathcal{D}$, the mappings $\boldsymbol{\sigma}, \boldsymbol{\tau}$ inducing the partitions $\{\tilde{\mathcal{V}}_i\}_{\mathbb{Z}_2}$ and $\{\tilde{\mathcal{V}}_i^{\dagger}\}_{\mathbb{Z}_2}$ of $\tilde{\mathcal{V}}$.

A vertex in $\hat{\mathcal{D}}$ we denote by $\boldsymbol{\mathcal{V}}_i$; $c \in \mathcal{C}$.

The structurally/directionally dual constructions, *geometrical co-source-minor*, $\hat{\mathcal{D}}^{\perp}$/*geometrical basin-minor*, $\hat{\mathcal{D}}^{\dagger}$ are introduced as expected.

2.  A *special Clifford walk*, $\mathcal{R}$, on $\mathcal{D}_{\gamma}$, with underlying discrete-time Markov chain $\left(X_t : t = 0,1,2,\ldots\right)$ — a random walk, $\mathcal{R}$, on $\mathcal{D}_{\gamma}$. This chain $\left(X_t\right)$ moves according to the transition map $\chi : V \to V$, defined by $v_j \mapsto \sum_i \left(\gamma_{ji} / \sum_{m \in N_i^{in}} \gamma_{mi}\right) v_i$. We write $\wp_{ij}$ for the $(i,j)$ entry of the stochastic matrix representing $\chi$. Associated to $\mathcal{R}$ is the endomorphism of free $\tilde{V}$-modules, $\boldsymbol{\chi} : \wp \mathcal{V} \to \wp \mathcal{V}$;

$$v_j \mapsto \sum_{i : i \neq j} \wp_{ij} \omega_{j \to i}^{\top} v_i + \wp_{jj} \omega_{j \to j} v_j,$$

where, for all $j \in \mathcal{I}$, $\omega_{j \to i} \in \mathcal{B}$.

**Proposition.** *There is an endomorphism* $\tilde{\chi} : {}_{N}V \to {}_{N}V$ *of projective (left)* $\tilde{V}$*-modules induced by* $\boldsymbol{\chi}$.

The restriction of $\mathcal{R}$ to $\mathcal{D}_{\gamma}|_{\mathcal{E}_c} := \left(\mathcal{D}|_{\mathcal{E}_c}, \gamma|_{\mathcal{E}_c}\right)$ is related to the *c-transition map*, $\chi_c$; $c \in \mathcal{C}$. We write $\left(\wp_{ij}^c\right)$ for the substochastic matrix representing each such map $\chi_c$. Then, if, for $i \in \mathcal{I}$, $\sum_j \wp_{ij}^c < \max_i \left(\sum_j \wp_{ij}^c\right)$, $v_i$ is a *deprived* vertex in $\mathcal{D}_{\gamma}|_{\mathcal{E}_c}$. Otherwise, we say that $v_i$ is *privileged* in $\mathcal{D}_{\gamma}|_{\mathcal{E}_c}$. In particular, if $\min_i\left(\sum_j \wp_{ij}^c\right) = \max_i\left(\sum_j \wp_{ij}^c\right)$, every vertex in $\mathcal{D}_{\gamma}|_{\mathcal{E}_c}$ is privileged.

For every $c \in \mathcal{C}$, we denote by $\nu^c$ ($\pi^c$) in $V$, with $\sum \nu_i^c \pi_i^c = 1$, the right (left) Perron vector of $\left(\wp_{ij}^c\right)$ whenever this vector exists. And we let $\pi$ in $V$, with $\sum \pi_i = 1$, be a unique left fixed point of $\chi$, namely the left Perron vector of $\left(\wp_{ij}\right)$ whenever this vector is defined.

REMARK: By directional duality, one defines a special Clifford walk, $\boldsymbol{\mathcal{R}}^{\uparrow}$, on $\boldsymbol{\mathcal{D}}_\gamma$, with underlying Markov chain $\left( X_i^{\uparrow} \right)$ directed forwards; in particular, $\left( X_i^{\uparrow} \right)$ is a random walk, $\mathcal{R}^{\uparrow}$, on $\mathcal{D}_\gamma$.

3. a subset $\boldsymbol{\mathcal{F}}$ of *spinorial flows*, $\boldsymbol{\varphi} : \mathcal{V} \to \tilde{\mathrm{V}}$, in $\left( \mathbb{O}^\mathcal{V}, \mathbb{C} \right)$, definable in terms of flows, $\varphi : \mathrm{V} \to \mathbb{R}_0$, in a subset $\mathcal{F}$ of $\left( \mathrm{V}, \mathrm{E} \right)$.

Each of these maps $\boldsymbol{\varphi}$ assign to every $v_i$ in $\mathcal{V}$ $\left( \left( v_i, v_j \right) \right.$ in $\left. \mathcal{E} \right)$ a unique element, $\boldsymbol{\varphi}\left( v_i \right)$ (resp., $\boldsymbol{\varphi}\left( v_i, v_j \right)$) in $\tilde{\mathrm{V}} \simeq \mathfrak{C}_{p,q}^0$ with nonnegative coefficients; $\varphi$ is introduced in an analogous way. We occasionally write $\boldsymbol{\varphi}_{ij}$ $\left( \varphi_{ij} \right)$ for $\boldsymbol{\varphi}\left( v_i, v_j \right)$ (resp., $\varphi\left( v_i, v_j \right)$).

Let us denote by $\square\, \boldsymbol{\varphi} : \mathcal{V} \to \tilde{\mathrm{V}}$ the map given by

$$v_i \mapsto \sum_{j \in \mathcal{N}_i^{\text{out}}} \boldsymbol{\varphi}_{ij}^{\tau} - \sum_{j \in \mathcal{N}_i^{\text{in}}} \boldsymbol{\varphi}_{ji}^{\tau}$$

or, equivalently, by $v_i \mapsto \sum_{j \in \mathcal{N}_i^{\text{out}}} \varphi_{ij} \omega_{i \to j}^{\tau} - \sum_{j \in \mathcal{N}_i^{\text{in}}} \varphi_{ji} \, \omega_{j \to i}^{\tau}$ ; $i \in \mathcal{I}$ .

DEFINITION (**mirror-super(a)symmetric flow**) A flow $\boldsymbol{\varphi}^* \in \mathcal{F}$ $\left( \varphi^* \in \mathcal{F} \right)$ is *mirror-supersymmetric* or, simply, *mirror-symmetric* in $\boldsymbol{\mathcal{D}}_\gamma$ (resp., in $\mathcal{D}_\gamma$) if, for every $i \in \mathcal{I}$ ,

$$\square\, \boldsymbol{\varphi}^*\left( v_i \right) \in \boldsymbol{\mathcal{O}}\left( \delta \right) \text{ and, } \textit{self-dually, } \square\, \boldsymbol{\varphi}^*\left( v_i \right) \in \boldsymbol{\mathcal{O}}_{p,q}^0 \tag{1}$$

with respect to the special quadratic algebra $\left( \tilde{\mathrm{V}}, \delta \right) \simeq \left( \mathfrak{C}_{p,q}^0, \delta \right)$ ; $p \neq 0$ . The case $p = 0$ furnishes the (super-)asymmetric counterpart.

By proposition I.1, a mirror-supersymmetric flow in $\boldsymbol{\mathcal{D}}_\gamma$ is definable for $\dim_\mathbb{R} \mathrm{X} = p + q \leq 5$ .

The composition $\delta \circ \square\, \boldsymbol{\varphi} : \mathcal{V} \to \tilde{\mathrm{V}} \to \tilde{\mathrm{V}}$ has a unique extension to the $\mathbb{R}$-homomorphism $\mathrm{V} \to \tilde{\mathrm{V}}$, and thereby also to the homomorphism of left $\tilde{\mathrm{V}}$-modules $_{\tilde{\mathrm{V}}} \mathrm{V} \to \tilde{\mathrm{V}}$ . This extension we also denote by $\delta \circ \square\, \boldsymbol{\varphi}$ . It follows by linearity:

**Proposition.** *If $\varphi^* \in \mathcal{F}$ is mirror-(a)symmetric in $\mathcal{D}_\gamma$ , then $\varphi^*$ is mirror-(a)symmetric in $\left( {}_{\tilde{\mathrm{V}}}\mathrm{V}, {}_{\tilde{\mathrm{V}}}\mathrm{E} \right)$ ; for every $\mathrm{v} \in {}_{\tilde{\mathrm{V}}}\mathrm{V}$ ,* $\delta \circ \square\, \boldsymbol{\varphi}^*\left( \mathrm{v} \right) = 0_{\tilde{\mathrm{V}}}$ .

The spinorial flow $\boldsymbol{\varphi}_\pi \in \mathcal{F}$ $\left( \varphi_\pi \in \mathcal{F} \right)$, given by $\left( v_j, v_i \right) \mapsto \pi_i \, \wp_{ij} \omega_{j \to i}$ (resp., $\left( v_j, v_i \right) \mapsto \pi_i \, \wp_{ij}$) is said to be *stochastic* in $\boldsymbol{\mathcal{D}}_\gamma$ (resp., in $\mathcal{D}_\gamma$). Because the image of the mapping $\mathcal{V} \to \mathbb{R}_0$ ; $v_i \mapsto \sum_{j \in \mathcal{N}_i^{\text{out}}} \varphi_{ij} - \sum_{j \in \mathcal{N}_i^{\text{in}}} \varphi_{ji}$ vanishes trivially for $\varphi = \varphi_\pi$ , $\varphi_\pi \in \mathcal{F}$ is always a stochastic circulation in $\mathcal{D}_\gamma$ , but not necessarily a mirror-(a)symmetric flow.

By a **c**-*flow* in $\boldsymbol{\mathcal{D}}_\gamma$ , we refer to the spinorial flow $\boldsymbol{\varphi}_{\pi^c ; c \in \mathcal{C}}^* \in \mathcal{F}$ $\left( \varphi_{\pi^c ; c \in \mathcal{C}}^* \in \mathcal{F} \right)$ given by $\left( v_j, v_i \right) \mapsto \pi_i^c \, \wp_{ij} \nu_j^c \, \omega_{j \to i}$ (resp., $\left( v_j, v_i \right) \mapsto \pi_i^c \, \wp_{ij} \nu_j^c$) for $\omega_{j \to i} = b_c$ , with $c \in \mathcal{C}$ . This flow we abbreviate as $\boldsymbol{\varphi}_\mathbf{c}$ (resp., $\varphi_\mathbf{c}$ ). If $\boldsymbol{\varphi}_\mathbf{c}^*$ is mirror-(a)symmetric in $\boldsymbol{\mathcal{D}}_\gamma$ , we also say $\boldsymbol{\varphi}_\mathbf{c}^*$ is a *mirror-**c**-flow* in $\boldsymbol{\mathcal{D}}_\gamma$ , since, in essence, the symmetric case is here not different from the asymmetric one.

This completes the description of the flow network $\boldsymbol{\mathcal{F}}$ .

**II.2.1** The following result is well-established in the literature.

**Lemma.** *(i) A stationary distribution of the Markov chain $\left(X_t\right)$ $\left(\left(X_t^{\top}\right)\right)$ on $\mathcal{V}$ is unique if and only if $\mathcal{D}$ has a repellor, $\mathcal{T}_{\mapsto s}$, with source $\left[v_s\right]$ (resp., an attractor, $\mathcal{T}_{\to b}$, with basin $\left[v_b\right]$), that is, if and only if the strong minor of $\mathcal{D}$ has a unique initial vertex $\left[v_s\right]$ (resp., a unique terminal vertex $\left[v_b\right]$).*

*(ii) If $\mathcal{D}$ has a repellor, $\mathcal{T}_{\mapsto s}$, with source $\left[v_s\right]$, $\pi_t = 0$ if and only if vertex $v_t$ is submissive to $\left[v_s\right]$.*

Condition *(i)* in this lemma is equivalent to convergence of the random walk $\mathcal{R}$ ($\mathcal{R}^{\top}$) on $\mathcal{D}_\gamma$ to its stationary distribution, the walk's mixing time being controllable by the Cheeger constant of $\mathcal{D}_\gamma$.

**II.2.2** Let us assume that $\mathcal{D}$ has a repellor, $\mathcal{T}_{\mapsto s}$, with source $\left[v_s\right] = \mathcal{V}$. By $\mathcal{H} \subset \mathcal{V}$ we denote a nonempty proper subset of $\mathcal{V}$, and by $\partial_{in}\mathcal{H}$ we refer to the in-boundary of $\mathcal{H}$ relative to the cut $\left\{\mathcal{H}, \overline{\mathcal{H}}\right\}$ in $\mathcal{D}_{\mathcal{V}}$, where $\overline{\mathcal{H}} = \mathcal{V} \setminus \mathcal{H}$. Assuming further that the stochastic flow $\varphi_\pi^*$ is mirror-(a)symmetric in $\mathcal{D}_\gamma$, we define the *volume*, $V\left(\mathcal{H}\right)$, of $\mathcal{H}$ by the expression $V\left(\mathcal{H}\right) := \sum_{v_i \in \mathcal{H}} \sum_{v_j \in \mathcal{V}} \varphi_\pi^*\left(v_j, v_i\right)$, while for the *area*, $A\left(\partial_{in}\mathcal{H}\right)$, of $\partial_{in}\mathcal{H}$ we write $A\left(\partial_{in}\mathcal{H}\right) := \sum_{v_i \in \mathcal{H}} \sum_{v_j \in \overline{\mathcal{H}}} \varphi_\pi^*\left(v_j, v_i\right)$.

Under the spin geometry $\mathcal{D}_\gamma$, the *Cheeger constant* of $\mathcal{D}_\gamma$ is given by

$$h := \inf_{\mathcal{H} \subset \mathcal{V}} \frac{A\left(\partial_{in}\mathcal{H}\right)}{\min\left\{V\left(\mathcal{H}\right), V\left(\overline{\mathcal{H}}\right)\right\}}. \tag{2}$$

**II.3.1 Lemma.** *The $\mathbf{c}$-flow $\varphi_{\mathbf{c}} \in \mathcal{F}$ exists in $\mathcal{D}_\gamma$ if, for every $c \in \mathcal{C}$, there exists a subset $\hat{\mathcal{V}}_c \subseteq \mathcal{V}$ of privileged vertices in $\mathcal{D}_\gamma \big| \mathcal{E}_c$, which is non-reachable from the subset of deprived vertices, and such that the induced subgraph $\mathcal{D} \big| \mathcal{E}_c \left\langle \hat{\mathcal{V}}_c \right\rangle$ has a repellor, $\mathcal{T}_{\mapsto c}$, with source $\left[v_c\right] \subseteq \mathcal{V}_c$.*

*Proof:* Let $\lambda_c$ denote the Perron root of $\left(\wp_{ij}^c\right)$. It is a known result [T.18] that $\lambda_c = \max_i \left(\sum_j \wp_{ij}^c\right)$ if and only if there exists a subset $\hat{\mathcal{V}}_c \subseteq \mathcal{V}$ of privileged vertices in $\mathcal{D}_\gamma \big| \mathcal{E}_c$, which is non-reachable from the subset of deprived vertices. If this is the case, $\left(\wp_{ij}^c\right)$ is reducible and permutation similar to a matrix with a unique ($\left|\hat{\mathcal{V}}_c\right|$-by-$\left|\hat{\mathcal{V}}_c\right|$) diagonal block whose Perron root is $\lambda_c$. By [T.19], $\lambda_c$ is algebraically simple if and only if $\mathcal{D} \big| \mathcal{E}_c \left\langle \hat{\mathcal{V}}_c \right\rangle$ has a repellor, $\mathcal{T}_{\mapsto c}$, with source $\left[v_c\right]$, and mutual reachability implies that $\left[v_c\right] \subseteq \mathcal{V}_c$. Then, since the geometric multiplicity of $\lambda_c$ as an eigenvalue of $\left(\wp_{ij}^c\right)$ is 1, there exist a unique element $\nu^c$ ($\pi^c$) in $\mathbb{V}$, with $\sum_i \nu_i^c \pi_i^c = 1$, comprising the right (resp., left) Perron vector of $\left(\wp_{ij}^c\right)$. Under this condition, the $\mathbf{c}$-flow $\varphi_{\mathbf{c}} \in \mathcal{F}$ ($\varphi_{\mathbf{c}} \in \mathcal{F}$) exists in $\mathcal{D}_\gamma$ (resp., in $\mathcal{D}_\gamma$) and, for every $i \in \mathcal{I}$, we trivially have $\square\varphi_{\mathbf{c}}\left(v_i\right) \in \mathcal{O}\left(\delta\right)$. $\blacksquare$

**II.3.2** Let us assume that the condition stated in II.3.1 is satisfied. We write $\mathcal{V}^c$ for $\left[v_c\right]$. By $\mathcal{H}^c \subset \mathcal{V}^c$ we denote a nonempty proper subset of $\mathcal{V}^c$, and by $\partial_{in}\mathcal{H}^c$ we refer to the in-boundary of $\mathcal{H}^c$ relative to the cut $\left\{\mathcal{H}^c, \overline{\mathcal{H}}^c\right\}$ in $\mathcal{D}\left\langle\mathcal{V}^c\right\rangle$, where $\overline{\mathcal{H}}^c = \mathcal{V}^c \setminus \mathcal{H}^c$. Similarly to I.2.2, assuming that $\varphi_{\mathbf{c}}^*$ is a mirror-$\mathbf{c}$-flow, in $\mathcal{D}_\gamma$, we define the volume, $V\left(\mathcal{H}^c\right)$, of $\mathcal{H}^c$ as

$V\left(\mathcal{H}^c\right) := \sum_{v_i \in \mathcal{H}^c} \sum_{v_j \in \mathcal{V}^c} \varphi_{\mathbf{c}}^*\left(v_j, v_i\right)$, while for the area, $A\left(\partial_{in}\mathcal{H}^c\right)$, of $\partial_{in}\mathcal{H}^c$ we write $A\left(\partial_{in}\mathcal{H}^c\right) := \sum_{v_i \in \mathcal{H}^c} \sum_{v_j \in \overline{\mathcal{H}}^c} \varphi_{\mathbf{c}}^*\left(v_j, v_i\right)$.

Under the spin geometry $\mathcal{D}_\gamma$, the *product Cheeger constant* of $\mathcal{D}_\gamma$ is given by

$$\hbar := \prod_c \inf_{\mathcal{H}^c \subset \mathcal{V}} \frac{A\left(\partial_{in}\mathcal{H}^c\right)}{\min\left\{V\left(\mathcal{H}^c\right), V\left(\bar{\mathcal{H}}^c\right)\right\}}. \qquad (3)$$

**II.4** REMARK: By directional duality, the assertions presented in II.2.2 and II.3.1-3.2 can be appropriately restated for a special Clifford walk, $\mathcal{R}^\uparrow$, on $\mathcal{D}_\gamma$. The same holds true for statements that are to follow.

**Proposition.** *Let $\mathcal{F}_*$ be a mirror-supersymmetric flow network. A distinction between $\mathcal{F}_*$ and $\mathcal{F}_*^\uparrow$ reflects a distinction between the 'future' and 'past' super-cones in $\mathcal{O}\left(\delta\right)$.*

**II.5** We denote by $\mathscr{O}$ a self-organizing (sub)cortical circuit, with orientation-preserving, rotational or Lorentzian dynamics.

**Axiom.** *There is a flow network $\mathcal{F}$ underlying $\mathscr{O}$.*

DEFINITION **(super(a)symmetry)** $\mathscr{O}$ is *supersymmetric* whenever its underlying flow network, $\mathcal{F}$, is, i.e., whenever $\left(\tilde{V}, \tilde{\vartheta}\right)$ is hyperbolic. It is *super-asymmetric* otherwise.

**Proposition 1.** *The mapping $\sigma : \tilde{\mathcal{V}} \to \mathbb{Z}_2$ in $\mathcal{D}_\gamma$ is a latent indicator of the neuron type (excitatory vs. inhibitory) and, therewith, of matter vs. antimatter in $\mathscr{O}$.*

**Proposition 2.** *The mapping $\tau : \tilde{\mathcal{V}} \to \mathbb{Z}_2$ in $\mathcal{D}_\gamma$ is a latent indicator of the inhibitory neuron type and, therewith, of super-(a)symmetry in $\mathscr{O}$.*

The experimental observations in [T.20] reinforce and are reinforced by proposition 2, by virtue of the observations in section III, suggesting the link: fast-spiking interneurons − reciprocity − supersymmetry − Lorentzian dynamics.

Rotational or Lorentzian co-dynamics may alternatively be observed in $\mathscr{O}$. As well, the direction of flow may alter.

**II.6 Proposition 1.** *The geometrical (source-)minor of $\mathcal{D}$ is a population-level structure.*

Certain parallels to the tenets of *dimensionality reduction* [T.21] can be drawn by the reader.

**Proposition 2.** *The spin geometry $\mathcal{D}_\gamma$ is a hybrid structure.*

**II.7 Proposition 1.** *Assume that $\mathcal{D}$ has a repellor, $\mathcal{T}_{\leftarrow_s}$, with source $\mathcal{V}$. Self-duality in (1) for the mirror-(a)symmetric stochastic flow $\varphi_\pi^* \in \mathcal{F}$ implies self-duality of the induced subdigraph $\mathcal{D}_\mathcal{V}$.*

*Proof:* While we do not attempt to provide a rigorous proof [T.22], we observe that condition (1) necessitates the existence of a self-dual (sub)digraph corresponding to the existence of spinor co-representations (as defined by the self-duality

$$\mathbf{Spin}^+\left(p, q\right) = \left\{x \in \mathfrak{C}_{p,q}^0 : \delta\left(x\right) = 1_{\mathfrak{C}_{p,q}}\right\},$$ for $p + q \leq 5$ ). Since the Cheeger constant in (2) is defined exclusively in terms of the source $\mathcal{V}$ of $\mathcal{D}$, this self-dual subdigraph must be $\mathcal{D}_\mathcal{V}$. ∎

**Proposition 2.** *Assume that condition* II.3.1 *is satisfied. Self-duality in* (1) *for the mirror-**c**-flow* $\varphi_c^* \in \mathcal{F}$ *implies self-duality of the induced subgraphs* $\mathcal{D}\left\langle \Psi^c \right\rangle$ ; $c \in \mathcal{C}$ .

*Proof:* This follows an argument similar to that of proposition 1, considering the form of the product Cheeger constant in (3). ∎

Self-duality in the above claims refers to a special case of *geometric duality* observed inter alia in Platonic graphs [T.23]. Consistent with the terminology in section I, we name this form of self-duality *Wittgenstein*- [T.1] or $\mathcal{GP}$-*duality*.

**II.8** Further references: [T.24-33].

# III. Existence Theorems

**III.1 Theorem.** *Let* $\tilde{V} \simeq \mathfrak{C}_{0,q}^0$ . *The spinorial flow* $\varphi_\pi^*$ *is mirror-superasymmetric in* $\mathcal{D}_\gamma$ *if and only if* $\mathcal{D}$ *has a repellor,* $\mathcal{T}_{\leftarrow s}$ , *with source* $\left[ v_s \right] \subseteq \mathcal{V}_s$ , $s \in \mathcal{C}$ , *such that the induced subgraph* $\mathcal{D}\left\langle \left[ v_s \right] \right\rangle$ *is* $\mathcal{GP}$-*dual.*

*Proof:* Necessity of a repellor, $\mathcal{T}_{\leftarrow s}$ , with source $\left[ v_s \right]$ in $\mathcal{D}$ is implied by lemma II.2.1. $\mathcal{GP}$-duality of $\mathcal{D}\left\langle \left[ v_s \right] \right\rangle$ follows from proposition II.7.1.

Let $x_{j;\varphi} := \sum_{i \in \mathcal{N}_j^{out}} \omega\left( v_j, \vec{e}_{ji} \right) \varphi\left( v_j, v_i \right) + \sum_{i \in \mathcal{N}_j^{in}} \omega\left( v_j, \vec{e}_{ji} \right) \varphi\left( v_i, v_j \right)$ . We write $x_{j;\pi}$ for $x_{j;\varphi_\pi^*}$ . For $\varphi_\pi^*$ to be mirror-asymmetric in $\mathcal{D}_\gamma$ , for all $j \in \mathcal{I}$ , $x_{j;\pi}$ must be in $\mathcal{O}\left( \delta \right)$ , with respect to the special quadratic algebra $\left( \tilde{V}, \delta \right)$ . Accordingly, the anisotropic quadratic form, $\tilde{\vartheta}$ , on $\tilde{V}$ (considered in its normalized form), must necessarily vanish at $x_{j;\pi}$ , i.e., for all $j \in \mathcal{I}$ ,

$$\sum_{c \in \mathcal{C}} \left( \sum_{i \in \mathcal{N}_j^{out(c)}} \varphi_\pi^*\left( v_j, v_i \right) - \sum_{i \in \mathcal{N}_j^{in(c)}} \varphi_\pi^*\left( v_i, v_j \right) \right)^2 = 0 . \tag{1}$$

Let $j \in \mathcal{I}_t$ , with $t \in \mathcal{C}$ . Since $\sum_{i \in \mathcal{N}_j^{out}} \pi_j \wp_{ij} = \pi_j = \sum_{i \in \mathcal{N}_j^{in}} \pi_j \wp_{ji}$ , and $\mathcal{N}_j^{out(c)} = \emptyset$ for all $c \neq t$ , (1) is satisfied if and only if

$$\pi_j \sum_{i \in \mathcal{N}_j^{in(c)}} \wp_{ji} = 0 , \text{ for all } c \neq t . \tag{2}$$

Clearly, (2) holds if and only if $\pi_j = 0$ and/or $\mathcal{N}_j^{in(c)} = \emptyset$ , for all $c \neq t$ ; i.e., by lemma II.2.1, if and only if $v_j$ is submissive and/or $\mathcal{N}_j^{in(c)} = \emptyset$ , for all $c \neq t$ . Since the source $\left[ v_s \right]$ is non-submissive, for all $v_{j_s} \in \left[ v_s \right]$ and $c \neq t$ , we must have $\mathcal{N}_j^{in(c)} = \emptyset$ ; i.e., $\left[ v_s \right] \subseteq \mathcal{V}_s$ ; $s \in \mathcal{C}$ . Then, for all $j \in \mathcal{I}$ , $\tilde{\vartheta}\left( x_{j;\pi} \right) = 0$ implies $\delta\left( x_{j;\pi} \right) = 0_{\tilde{V}}$ . Hence, $x_{j;\pi} \in \mathcal{O}\left( \delta \right)$ . ∎

**III.2** We now turn to examine certain cases where $\left( \tilde{V}, \tilde{\vartheta} \right)$ is a hyperbolic quadratic space. It is intuitively clear that, in such cases, for the flow $\varphi_\pi^*$ to be mirror-supersymmetric in $\mathcal{D}_\gamma$ , the source need not be constrained to lie in a single deep cluster in $\mathcal{D}$ .

**Theorem.** Let $\tilde{V} \simeq \mathfrak{C}^0_{1,3}$ . The spinorial flow $\varphi^*_\pi$ is mirror-supersymmetric in $\mathcal{D}_\gamma$ if and only if $\mathcal{D}$ has a repellor, $\mathcal{T}_{\rightarrow s}$ , with source $\mathcal{V} \subseteq \mathcal{V}_1 \cup \ldots \cup \mathcal{V}_8$ , such that, up to a suitable permutation of the index set $\mathcal{C} = \mathcal{I}$ , the following conditions are satisfied

**(i)** the induced subdigraph $\mathcal{D}_{\boldsymbol{v}}$ is $\mathcal{H}$-dual;

**(ii)** if $\tilde{\mathcal{V}}$ has cardinality $|\tilde{\boldsymbol{\mathcal{V}}}| > 2$ , the geometrical source-minor, $\tilde{\boldsymbol{\mathcal{D}}}$ , of $\mathcal{D}$ has bipartition $\left( \tilde{\boldsymbol{\mathcal{V}}}^\dagger_0, \tilde{\boldsymbol{\mathcal{V}}}^\dagger_1 \right)$ ;

**(iii)** if $\tilde{\boldsymbol{\mathcal{V}}}_0$ has cardinality $\left| \tilde{\boldsymbol{\mathcal{V}}}_0 \right| = 2$ , the induced subdigraph $\tilde{\boldsymbol{\mathcal{D}}}\left\langle \tilde{\boldsymbol{\mathcal{V}}}_0 \right\rangle$ is disconnected;

**(iv)** if $\left\{ \boldsymbol{\mathcal{V}}_i, \boldsymbol{\mathcal{V}}_j \right\} \in \tilde{\mathcal{V}}^\dagger_1$ , the induced subdigraph $\tilde{\boldsymbol{\mathcal{D}}}\left\langle \left\{ \boldsymbol{\mathcal{V}}_i, \boldsymbol{\mathcal{V}}_j \right\} \right\rangle$ is disconnected;

**(v)** for every $v_j \in \boldsymbol{\mathcal{V}}$ , with $j \in \mathcal{I}_s$ and $s \in \mathcal{C}^\dagger_i$ , $\partial_{in} v_j \subseteq \mathcal{V}_s \cup \mathcal{V}_t$ , for some $t \in \mathcal{C}^\dagger_{i+1}$ ,

the indices in (iv) following the convention given in the proof below.

*Proof:* Necessity of a repellor, $\mathcal{T}_{\rightarrow s}$ , with source $\left[ v_s \right] = \boldsymbol{\mathcal{V}}$ in $\mathcal{D}$ is implied by lemma II.2.1. $\mathcal{H}$-duality of $\mathcal{D}_{\boldsymbol{v}}$ follows from proposition II.7.1.

We let $x = a_1 \cdot 1_{\mathfrak{C}^0_{1,3}} + a_2 e_1 e_2 + \ldots + a_i e_{i_1} e_{i_2} + \ldots + a_7 e_3 e_4 + a_8\, e_1 e_2 e_3 e_4$ , with $a_i \in \mathbb{R}$ , be an arbitrary element in $\tilde{V}$ , where we identify $\tilde{\mathcal{V}}$ with an orthonormal basis for $\mathfrak{C}^0_{1,3}$ . The quadratic norm $\delta$ on $\tilde{V}$ is given by $\delta(x) = \tilde{\vartheta}(x) \cdot 1_{\mathfrak{C}^0_{1,3}} + 2\left( a_1 a_8 - a_2 a_7 \right) e_1 e_2 e_3 e_4$ .

For $i \in \mathbb{Z}_2$ and $t \in \mathcal{C}^\dagger_i$ , let $c_1 \neq \ldots \neq c_d \neq \ldots \neq c_{u-1} \neq t$ , with $c_d \in \mathcal{C}^\dagger_i$ , and let $f_1 \neq \ldots \neq f_d \neq \ldots \neq f_u$ , with $f_d \in \mathcal{C}^\dagger_{i+1}$ .

Let $x_{j;\varphi} := \sum_{i \in \mathcal{N}^{out}_j} \omega\left( v_j, \overleftarrow{e}_{ji} \right) \varphi\left( v_j, v_i \right) + \sum_{i \in \mathcal{N}^{in}_j} \omega\left( v_j, \overleftarrow{e}_{ji} \right) \varphi\left( v_i, v_j \right)$ . We write $x_{j;\pi}$ for $x_{j;\varphi^*_\pi}$ . For $\varphi^*_\pi$ to be mirror-symmetric in $\mathcal{D}_\gamma$ , for all $j \in \mathcal{I}$ , $x_{j;\pi}$ must be in $\mathcal{O}(\delta)$ , with respect to the special quadratic algebra $\left( \tilde{V}, \delta \right)$ . Accordingly, the isotropic, non-degenerate quadratic form, $\tilde{\vartheta}$ , on $\tilde{V}$ (considered in its normalized form), must necessarily vanish at $x_{j;\pi}$ , i.e., for all $j \in \mathcal{I}$ ,

$$\tilde{\vartheta}\left( x_{j;\pi} \right) = 0 \ . \tag{1}$$

As well, for all $j \in \mathcal{I}$ , the following condition must hold:

$$a_1 a_8 - a_2 a_7 = 0 \ . \tag{2}$$

Since $\sum_{i \in \mathcal{N}^{out}_j} \pi_j \wp_{ij} = \pi_j = \sum_{i \in \mathcal{N}^{in}_j} \pi_j \wp_{ji}$ , the former condition, (1), forces an homogeneous polynomial of degree 2 in $\mathbb{R}\left[ \pi_j \sum_{i \in \mathcal{N}^{in(c_1)}_j} \wp_{ji}, \ldots, \pi_j \sum_{i \in \mathcal{N}^{in(c_d)}_j} \wp_{ji}, \ldots, \pi_j \sum_{i \in \mathcal{N}^{in(c_{u-1})}_j} \wp_{ji}, \ \pi_j \sum_{i \in \mathcal{N}^{in(f_1)}_j} \wp_{ji}, \ldots, \pi_j \sum_{i \in \mathcal{N}^{in(f_d)}_j} \wp_{ji}, \ldots, \pi_j \sum_{i \in \mathcal{N}^{in(f_u)}_j} \wp_{ji} \right]$ to be zero, for all $j \in \mathcal{I}_t$ and $t \in \mathcal{C}^\dagger_i$ ; $i \in \mathbb{Z}_2$ . In particular, either $v_j$ must be submissive (by lemma II.2.1) and/or we must have: $\mathcal{N}^{in(c_d)}_j = \emptyset$ , for all $c_d \in \mathcal{C}^\dagger_i$ , with $c_d \neq t$ , *and* $\mathcal{N}^{in(f_d)}_j = \emptyset$ , for all but at most some $f_d \in \mathcal{C}^\dagger_{i+1}$ . This implies $\boldsymbol{\mathcal{V}} \subseteq \mathcal{V}_1 \cup \ldots \cup \mathcal{V}_k$ , *(ii)* and *(v)* being necessarily satisfied.

With respect to the latter condition, (2), first observe that, by the $\mathbb{Z}_2$-graded structure of $\tilde{V}$ as an $\mathbb{R}$-module, $\mathcal{V}_1, \mathcal{V}_8 \in \bar{\mathcal{V}}_0$, while $\mathcal{V}_2, \mathcal{V}_7 \in \bar{\mathcal{V}}_1$. In addition, if $\mathcal{V}_1, \mathcal{V}_i \in \bar{\mathcal{V}}_i^\dagger$, then $\mathcal{V}_2, \mathcal{V}_8 \in \bar{\mathcal{V}}_{i+1}^\dagger$; $i \in \mathbb{Z}_2$. Let $j \in \mathcal{I}_t$, with $t \neq 1, 2, 7, 8$. By the former condition, $a_1 a_8 - a_2 a_7 = \pi_j^2 \Big( \sum_{i \in \mathcal{N}_j^{in(1)}} \wp_{ji} \sum_{i \in \mathcal{N}_j^{in(8)}} \wp_{ji} - \sum_{i \in \mathcal{N}_j^{in(2)}} \wp_{ji} \sum_{i \in \mathcal{N}_j^{in(7)}} \wp_{ji} \Big)$ vanishes. Hence, $x_{j;\pi} \in \mathcal{O}\big(\delta\big)$. Now let $j \in \mathcal{I}_t$, with $t = 8 \in \mathcal{C}_0$ and $t \in \mathcal{C}_i^\dagger$; $i \in \mathbb{Z}_2$. By the former condition, $a_2 a_7 = \pi_j^2 \sum_{i \in \mathcal{N}_j^{in(2)}} \wp_{ji} \sum_{i \in \mathcal{N}_j^{in(7)}} \wp_{ji} = 0$. Thus, (2) reduces to

$$a_1 a_8 = 0 .$$

Obviously, if $\pi_j = 0$ or if $\mathcal{N}_j^{in(1)} = \emptyset$, $a_1 a_8 = -\pi_j \sum_{i \in \mathcal{N}_j^{in(1)}} \wp_{ji} \Big( \sum_{i \in \mathcal{N}_j^{in(8)}} \pi_i \wp_{ij} - \pi_j \sum_{i \in \mathcal{N}_j^{in(8)}} \wp_{ji} \Big) = 0$. If on the other hand, $\pi_j \neq 0$, $\mathcal{N}_j^{in(c_d)} = \emptyset$, for all $c_d \in \mathcal{C}_i^\dagger$, with $c_d \neq t$, and $\mathcal{N}_j^{in(f_d)} = \emptyset$ for all $f_d \in \mathcal{C}_{i+1}^\dagger$, with $f_d \neq 1$, then, by the former condition, $a_1 a_8 = \pi_j^2 \Big( \sum_{i \in \mathcal{N}_j^{in(1)}} \wp_{ji} \Big)^2$, which vanishes if and only if $\mathcal{N}_j^{in(1)} = \emptyset$. By an analogous argument, we find that if $j \in \mathcal{I}_t$, with $t = 1 \in \mathcal{C}_0$, then, whenever $\pi_j \neq 0$, we must necessarily have $\mathcal{N}_j^{in(8)} = \emptyset$. Hence, if $\tilde{\mathcal{V}}_0$ has cardinality $\big| \tilde{\mathcal{V}}_0 \big| = 2$, the induced subdigraph $\tilde{\mathcal{D}} \big\langle \tilde{\mathcal{V}}_0 \big\rangle$ must be disconnected, i.e., (iii) must be true. If we consider the case where $j \in \mathcal{I}_t$, with $t = 2, 7 \in \mathcal{C}_1$, we obtain (iv) by a similar argument. ∎

**Corollary 1.** Let $\tilde{V} \simeq \mathfrak{C}_{1,3}^0$. Assume the spinorial flow $\varphi_\pi^*$ is mirror-supersymmetric in $\mathcal{D}_\gamma$. If $\tilde{\mathcal{V}}$ has cardinality $\big| \tilde{\mathcal{V}} \big| = 8$, the geometrical source-minor, $\tilde{\mathcal{D}}$, of $\mathcal{D}$ is (1,3)-partite; its Lorentz partition is exactly $\big( \tilde{\mathcal{V}}_0, \cdot \big)$.

**Corollary 2.** Let $\tilde{V} \simeq \mathfrak{C}_{1,2}^0$. The spinorial flow $\varphi_\pi^*$ is mirror-supersymmetric in $\mathcal{D}_\gamma$ if and only if $\mathcal{D}$ has a repellor, $\mathcal{T}_{\hookleftarrow s}$, with source $\mathcal{V} \subseteq \mathcal{V}_1 \cup \ldots \cup \mathcal{V}_4$, such that the following conditions are satisfied

**(i)** the induced subdigraph $\mathcal{D}_{\mathcal{V}}$ is $\mathcal{GP}$-dual.

**(ii)** if $\tilde{\mathcal{V}}$ has cardinality $\big| \tilde{\mathcal{V}} \big| > 2$, the geometrical source-minor, $\tilde{\mathcal{D}}$, of $\mathcal{D}$ has bipartition $\big( \tilde{\mathcal{V}}_0^\dagger, \tilde{\mathcal{V}}_1^\dagger \big)$;

**(iii)** for every $v_j \in \mathcal{V}$, with $j \in \mathcal{I}_s$ and $s \in \mathcal{C}_i^\dagger$, $\partial_{in} v_j \subseteq \mathcal{V}_s \cup \mathcal{V}_t$, for some $t \in \mathcal{C}_{i+1}^\dagger$.

**Corollary 2.1.** Let $\tilde{V} \simeq \mathfrak{C}_{1,2}^0$. Assume the spinorial flow $\varphi_\pi^*$ is mirror-supersymmetric in $\mathcal{D}_\gamma$. If $\tilde{\mathcal{V}}$ has cardinality $\big| \tilde{\mathcal{V}} \big| = 4$, the geometrical source-minor, $\tilde{\mathcal{D}}$, of $\mathcal{D}$ is (1,2)-partite; its Lorentz partition is exactly $\big( \tilde{\mathcal{V}}_0, \cdot \big)$.

**Corollary 3.** Let $\tilde{V} \simeq \mathfrak{C}_{1,1}^0$. The spinorial flow $\varphi_\pi^*$ is mirror-supersymmetric in $\mathcal{D}_\gamma$ if and only if $\mathcal{D}$ has a repellor, $\mathcal{T}_{\hookleftarrow s}$, with source $\mathcal{V} \subseteq \mathcal{V}_1 \cup \mathcal{V}_2$, such that the induced subdigraph $\mathcal{D}_{\mathcal{V}}$ is $\mathcal{GP}$-dual.

**III.3 Theorem.** *Let* $\tilde{V} \simeq \mathfrak{C}_{1,4}^0$ *. The spinorial flow* $\boldsymbol{\varphi}_\pi^*$ *is mirror-supersymmetric in* $\boldsymbol{\mathcal{D}}_\gamma$ *if and only if* $\boldsymbol{\mathcal{D}}$ *has a repellor,* $\mathcal{T}_{\to s}$ *, with source* $\boldsymbol{\mathcal{V}} \subseteq \mathcal{V}_1 \cup \ldots \cup \mathcal{V}_{16}$ *, such that, up to a suitable permutation of the index set* $\mathcal{C} = \mathcal{I}$ *, the following conditions are satisfied*

**(i)** *the induced subdigraph* $\mathcal{D}_{\boldsymbol{v}}$ *is* $\mathscr{W}$ *-dual;*

**(ii)** *if* $\tilde{\boldsymbol{\mathcal{V}}}$ *has cardinality* $\left|\tilde{\boldsymbol{\mathcal{V}}}\right| > 2$ *, the geometrical source-minor,* $\tilde{\boldsymbol{\mathcal{D}}}$ *, of* $\boldsymbol{\mathcal{D}}$ *has bipartition* $\left(\tilde{\boldsymbol{\mathcal{V}}}_0^\dagger, \tilde{\boldsymbol{\mathcal{V}}}_1^\dagger\right)$ *;*

**(iii)** *if* $\left\{\mathcal{V}_i, \mathcal{V}_c\right\} \in \tilde{\mathcal{V}}_0$ *, with* $c = 12, 13, 14$ *, the induced subdigraph* $\tilde{\boldsymbol{\mathcal{D}}}\left\langle\left\{\mathcal{V}_i, \mathcal{V}_c\right\}\right\rangle$ *is disconnected;*

**(iv)** *if* $\left\{\mathcal{V}_c, \mathcal{V}_{15}\right\} \in \tilde{\boldsymbol{\mathcal{V}}}$ *, with* $c = 6, 7, 8$ *, the induced subdigraph* $\tilde{\boldsymbol{\mathcal{D}}}\left\langle\left\{\mathcal{V}_c, \mathcal{V}_{15}\right\}\right\rangle$ *is disconnected;*

**(v)** *if* $\left\{\mathcal{V}_2, \mathcal{V}_{16}\right\} \in \tilde{\boldsymbol{\mathcal{V}}}$ *the induced subdigraph* $\tilde{\boldsymbol{\mathcal{D}}}\left\langle\left\{\mathcal{V}_2, \mathcal{V}_{16}\right\}\right\rangle$ *is disconnected;*

**(vi)** *for every* $v_j \in \boldsymbol{\mathcal{V}}$ *, with* $j \in \mathcal{I}_s$ *and* $s \in \mathcal{C}_i^\dagger$ *,* $\partial_m v_j \subseteq \mathcal{V}_s \cup \mathcal{V}_t$ *, for some* $t \in \mathcal{C}_{i+1}^\dagger$ *,*

*the indices in (iii)-(v) following the convention given in the proof below.*

*Proof:* The proof follows an argument analogous to that in theorem III.2, by letting

$x = a_1 \cdot 1_{\mathfrak{C}_{1,4}^0} + a_2 e_1 e_2 + a_3 e_1 e_3 + a_4 e_1 e_4 + a_5 e_1 e_5 + a_6 e_2 e_3 + a_7 e_2 e_4 + a_8 e_2 e_5 + a_9 e_3 e_4 + a_{10} e_3 e_5 + a_{11} e_4 e_5 +$

$+ a_{12} e_1 e_2 e_3 e_4 + a_{13} e_1 e_2 e_4 e_5 + a_{14} e_1 e_3 e_4 e_5 + a_{15} e_1 e_2 e_3 e_5 + a_{16} e_2 e_3 e_4 e_5$ *, with* $a_i \in \mathbb{R}$ *, be an arbitrary element in* $\tilde{V}$ *, where we identify* $\tilde{\mathcal{V}}$ *with an orthonormal basis for* $\mathfrak{C}_{1,4}^0$ *.* ∎

**Corollary.** *Let* $\tilde{V} \simeq \mathfrak{C}_{1,4}^0$ *. Assume the spinorial flow* $\boldsymbol{\varphi}_\pi^*$ *is mirror-supersymmetric in* $\boldsymbol{\mathcal{D}}_\gamma$ *. If* $\tilde{\boldsymbol{\mathcal{V}}}$ *has cardinality* $\left|\tilde{\boldsymbol{\mathcal{V}}}\right| = 16$ *, the geometrical source-minor,* $\tilde{\boldsymbol{\mathcal{D}}}$ *, of* $\boldsymbol{\mathcal{D}}$ *is (1,4)-partite.*

**III.4 Lemma.** *The spinorial flow* $\boldsymbol{\varphi}_c^*$ *is a mirror-*$\mathbf{c}$*-flow in* $\boldsymbol{\mathcal{D}}_\gamma$ *if condition II.3.1 is satisfied and the induced subdigraphs* $\mathcal{D}\left\langle\boldsymbol{\mathcal{V}}^c\right\rangle$ *, with* $c \in \mathcal{C}$ *, are* $\mathscr{W}$ *-dual.*

*Proof:* See II.3.1 and proposition II.7.2. ∎

**III.5 Proposition 1.** *By functional duality, one obtains the co-flow network* $\mathcal{F}^\perp$ *. Statements structurally dual to* II.2 -II.4, II.7 *and to those in* III.1-III.4 *hold true for the co-flow network* $\mathcal{F}^\perp$ *.*

**Proposition 2.** *Assume that the flow* $\boldsymbol{\varphi}_\pi^*$ *in* $\mathcal{F}_*$ *and the co-flow* $\boldsymbol{\varphi}_\pi^{*\perp}$ *in* $\mathcal{F}_*^\perp$ *are mirror-super(a)symmetric in* $\boldsymbol{\mathcal{D}}_\gamma$ *, and/or that the flow* $\boldsymbol{\varphi}_c^*$ *in* $\mathcal{F}_*$ *and the co-flow* $\boldsymbol{\varphi}_c^{*\perp}$ *in* $\mathcal{F}_*^\perp$ *are, respectively, a mirror-*$\mathbf{c}$*-flow and a mirror-*$\mathbf{c}$*-co-flow in* $\boldsymbol{\mathcal{D}}_\gamma$ *. Then*

$$\mathcal{F}_* = \mathcal{F}_*^\perp \,.$$

*That is, the flow network* $\mathcal{F}_*$ *is hyper-self-dual to the co-flow network* $\mathcal{F}_*^\perp$ *, under hyper-mirror super(a)symmetry.*

This form of self-duality we call *unifying* or $\mathcal{U}$-*duality*.

**III.6 Proposition.** *Another type of hyper-self-duality is feasible − the directional self-duality* $\mathcal{F}_* = \mathcal{F}_*^{\uparrow}$ , *under hyper-mirror super(a)symmetry. In fact, we can have*

$$\mathcal{F}_* = \mathcal{F}_*^{\uparrow} = \mathcal{F}_*^{\perp} = \mathcal{F}_*^{\perp\uparrow}\,.$$

## IV.  The Game On $\mathcal{F}$

**IV.1** In strategic (normal) form, a *$k$-person non-cooperative game on* $\mathcal{F}$ (or $\mathcal{F}$*-game*) is represented by an ordered triple $\left(\boldsymbol{\mathcal{P}}, \boldsymbol{\mathcal{S}}, \boldsymbol{\alpha}\right)$, where $\boldsymbol{\mathcal{P}} := \left(\mathcal{P}_c\right)_{\mathcal{C}}$ denotes an indexed family of sets − the *players*. Player $\mathcal{P}_c$ is associated to a family, $\mathcal{S}_c := \left(\mathcal{S}_{cj}\right)_{j \in \mathcal{J}_c}$, of *pure strategies*, with finite index set $\mathcal{J}_c$, so that $\boldsymbol{\mathcal{S}} := \left(\mathcal{S}_c\right)_{\mathcal{C}}$; and to a real-valued *utility* (or *payoff*) *function*, $\alpha_c : \times_c \left\{\mathcal{S}_{cj}\right\}_{j \in \mathcal{J}_c} \to \mathbb{R}$, such that $\boldsymbol{\alpha} := \left(\alpha_c\right)_{\mathcal{C}}$. The partition, $\left\{\mathcal{C}_i\right\}_{\mathbb{Z}_2}$, of the index set $\mathcal{C}$, which by Axiom II.1.1 is associated to the graded super-space structure of $\tilde{V} \simeq \mathfrak{C}_{p,q}^0$, defines a unique partition $\left\{\boldsymbol{\mathcal{P}}_i\right\}_{\mathbb{Z}_2}$ of the image set of $\boldsymbol{\mathcal{P}}$, according to whether player $\mathcal{P}_c$ is *proactive* or *counteractive*. On the other hand, the partition $\left\{\mathcal{C}_i^{\dagger}\right\}_{\mathbb{Z}_2}$ associated to the structure of $\left(\tilde{V}, \tilde{\vartheta}\right)$ as a nondegenerate real quadratic space gives rise to a distinct partition $\left\{\boldsymbol{\mathcal{P}}_i^{\dagger}\right\}_{\mathbb{Z}_2}$ of the image set of $\boldsymbol{\mathcal{P}}$, intimating the super(a)symmetry of the game.

A *mixed strategy*, $s_c$, of player $\mathcal{P}_c$ is a point in a simplex whose set of vertices is $\left\{\mathcal{S}_{cj}\right\}_{j \in \mathcal{J}_c}$, and thereby an element of a convex subset of an $\mathbb{R}$ - linear space, $S_c$, with a basis identified with the image set of $\mathcal{S}_c$. In turn, the multilinear map $\bar{\alpha}_c : \times_c S_c \to \mathbb{R}$ is the unique extension of the utility function $\alpha_c$ to the set $\times_c S_c$, so that $\bar{\alpha}_c\left(s_1, s_2, \ldots, s_k\right) = \hat{\alpha}_c\left(s_1 \otimes s_2 \otimes \ldots \otimes s_k\right)$, $\hat{\alpha}_c$ being the corresponding linear map $\otimes_c S_c \to \mathbb{R}$ .

Given a $k$-tuple, $\boldsymbol{s} = \left(s_c\right)_{\mathcal{C}}$, of mixed strategies in $\times_c S_c$, we let $\left(\boldsymbol{s}; t_c\right) := \left(s_1, s_2, \ldots, s_{c-1}, t_c, s_{c+1}, \ldots, s_k\right)$. A $k$-tuple, $\boldsymbol{s}^* = \left(s_c^*\right)_{\mathcal{C}}$, of mixed strategies is a *Nash equilibrium of an* $\mathcal{F}$*-game* if each player's mixed strategy is optimal against those of the other players, i.e., if for every $c \in \mathcal{C}$ :

$$\bar{\alpha}_c\left(\boldsymbol{s}^*\right) = \max_{s_c} \left[\bar{\alpha}_c\left(\boldsymbol{s}^*; s_c\right)\right].$$

At first, we may let player $\mathcal{P}_c$ be identified with deep cluster $\mathcal{V}_c$ in $\mathcal{D}$, and the associated family, $\mathcal{S}_c$, of pure strategies be the indexed family, $\bar{\mathcal{E}}_c := \left(\mathcal{E}_{cj}\right)_{j \in \mathcal{J}_c}$, of $c$-relations on $\mathcal{V}$ subject to Axiom II.1.2. Therewith, the utility function $\alpha_c$ may be identified with the map $\varepsilon_c : \times_c \left\{\mathcal{E}_{cj}\right\}_{j \in \mathcal{J}_c} \to \mathbb{R}$ (extended $\mathbb{R}$-linearly).

We define $\mathcal{G}\left(\mathcal{F}\right) := \left(\boldsymbol{\mathcal{V}}, \boldsymbol{\mathcal{E}}, \boldsymbol{\varepsilon}\right)$, where $\boldsymbol{\mathcal{V}} := \left(\mathcal{V}_c\right)_{\mathcal{C}}$, $\boldsymbol{\mathcal{E}} := \left(\bar{\mathcal{E}}_c\right)_{\mathcal{C}}$, and $\boldsymbol{\varepsilon} := \left(\varepsilon_c\right)_{\mathcal{C}}$. And we abbreviate $\mathcal{G}\left(\mathcal{F}\right)$ by $\mathcal{G}_{\mathcal{F}}$ .

REMARK: When introducing the form of a pure strategy in $\mathcal{G}_{\mathcal{F}}$, one may, in addition to $\mathcal{E}_{c_j}$, consider the map $\gamma\big|_{\mathcal{E}_{c_j}}$, therewith indicating that each player exercises control over the strength of the corresponding $c$-relations on $\mathcal{V}$. For simplicity of notation, we do not herein explicitly do so, but instead assume that such control is always implied.

**IV.2** For us, the structure of $\mathcal{G}_{\mathcal{F}}$ will bear similarities to the structure of a *prisoner's*- or *commons-dilemma* game, the *feasible and individually rational* payoffs being determined by conditions for which the spinorial flow $\varphi_{\pi}^* \in \mathcal{F}$ and/or the spinorial flow $\varphi_{\epsilon}^* \in \mathcal{F}$ are mirror-(a)symmetric in $\mathcal{D}_{\gamma}$. Pre-play communication and negotiations will be allowed. It goes without saying that $\mathcal{G}_{\mathcal{F}}$ is super-(a)symmetric whenever $\mathcal{F}$ is.

By proposition I.1

**Proposition.** *Let* $\tilde{V}_{\mathcal{F}} \simeq \mathfrak{C}_{1,5}^0$. *The dilemma game,* $\mathcal{G}_{\mathcal{F}}$, *on* $\mathcal{F}$ *restricts to a dilemma game,* $\mathcal{G}_{\hat{\mathcal{F}}}$, *on a flow subnetwork,* $\hat{\mathcal{F}}$, *of* $\mathcal{F}$, *with* $\tilde{V}_{\hat{\mathcal{F}}} \simeq \mathfrak{C}_{1,4}^0$.

Certain pure strategies in $\mathcal{G}_{\mathcal{F}}$ characterize a player in special ways.

DEFINITION (**self-reliant player**) We say that player $\mathcal{V}_c$ is *self-reliant* in $\mathcal{G}_{\mathcal{F}}$ whenever $\mathcal{V}_c$ adopts a (pure) strategy in $\bar{\mathcal{E}}_c$ respecting condition II.3.1, for $c \in \mathcal{C}$, while the induced subgraph $\mathcal{D}\langle \mathcal{V} \rangle$ is $\mathbb{OP}$-dual.

A *'cooperate vigorously'* (pure) strategy in $\bar{\mathcal{E}}_c$ is one where player $\mathcal{V}_c$ exercises control over the formation of a repellor in $\mathcal{D}$ with $\mathcal{V} \cap \mathcal{V}_c \neq \emptyset$, while being self-reliant. If $\tilde{V} \simeq \mathfrak{C}_{0,q}^0$, we require a strategy of this form to exhibit $\mathbb{OP}$-duality, in the sense of theorem III.1. If $\tilde{V} \simeq \mathfrak{C}_{1,4}^0$, $\tilde{V} \simeq \mathfrak{C}_{1,3}^0$ or $\tilde{V} \simeq \mathfrak{C}_{1,2}^0$, we require a strategy of this form to obey conditions *(i) - (vi)* in theorem III.3, *(i) - (v)* in theorem III.2, or *(i) - (iii)* in corollary II.2.2, respectively, under pre-play communication. We call a player $\mathcal{V}_c$ *submissive* in $\mathcal{G}_{\mathcal{F}}$ if $\mathcal{V}_c$ does not object to the formation of a repellor in $\mathcal{D}$, but does not exercise control over this structure either, i.e., $\mathcal{V} \cap \mathcal{V}_c = \emptyset$. Clearly, if this is the case, every vertex in $\mathcal{V}_c$ is submissive.

DEFINITION (**unconditionally submissive player**) A player, $\mathcal{V}_c$, is *unconditionally submissive* in $\mathcal{G}_{\mathcal{F}}$ if $\mathcal{V}_c$ is submissive without being self-reliant.

The *temptation*, player $\mathcal{V}_c$ is prone to, is to avoid contributing to the formation of a repellor in $\mathcal{D}$ in any way, while being self-reliant. Succumbing to such a temptation defines a *'defect'* (pure) strategy in $\bar{\mathcal{E}}_c$.

REMARK: In a sense, an unconditionally submissive player in $\mathcal{G}_{\mathcal{F}}$ is a player who cooperates unconditionally, while a player who submits with self-reliance is a player who, while cooperating, may be unwittingly defecting.

**IV.3** We are now led to adopt a Simonian hypothesis by drawing a parallel between a classification of certain pure strategies in $\mathcal{G}_{\mathcal{F}}$ and the *aspiration levels* in satisficing (as opposed to optimizing) approaches proposed within the theories of *bounded rationality*. In this regard, any two pure strategies of a player in $\mathcal{G}_{\mathcal{F}}$, pertaining to the same class (*level*), may be perceived as yielding approximately the same payoff, while *variation* − as reflected in the number of alternative $c$-relations falling under a single pure-strategy class − is not being compromised.

**Axiom.** *There is a reduced dilemma game,* $\tilde{G}_{\mathcal{F}}$*, of* $G_{\mathcal{F}}$*, wherein pure strategies are identified with the following homonymous levels of strategies in* $G_{\mathcal{F}}$*:* **(i)** *'cooperate vigorously' strategies;* **(ii)** *'submit with self-reliance' strategies;* **(iii)** *'submit unconditionally' strategies;* **(iv)** *'defect' strategies.*

DEFINITION (**Nash-Simon equilibrium**) A Nash equilibrium of the reduced dilemma game, $\tilde{G}_{\mathcal{F}}$, of $G_{\mathcal{F}}$ is a *Nash-Simon equilibrium* of $G_{\mathcal{F}}$.

DEFINITION (**Nash-Simon dimensionality**) Let $\tilde{S}^{*}$ denote the set of Nash-Simon equilibria of the dilemma game $G_{\mathcal{F}}$ and consider the function $\mathscr{I} : \tilde{S}^{*} \to \mathcal{C} \cup \{0\}$, given by $\tilde{s}^{*} \mapsto d$, $d$ denoting the number of 'cooperate vigorously' strategy-levels in $\tilde{s}^{*} \in \tilde{S}^{*}$. We say that a Nash-Simon equilibrium $\tilde{s}^{*} \in \tilde{S}^{*}$ of $G_{\mathcal{F}}$ has *(Nash-Simon) dimensionality* $\mathscr{I}\left(\tilde{s}^{*}\right)$.

**IV.4** Every finite non-cooperative game has a symmetric Nash equilibrium. Efficient outcomes in the reduced dilemma game, $\tilde{G}_{\mathcal{F}}$, of $G_{\mathcal{F}}$ are also achievable, the corresponding Nash equilibria being attainable in pure strategies. In fact, the following can be readily verified.

**Theorem 1.** *Let* $\tilde{V} \simeq \mathfrak{C}_{0,q}^{0}$*. A k-tuple of 'defect' strategy-levels is a symmetric Nash-Simon equilibrium of* $G_{\mathcal{F}}$*. A k-tuple with exactly one 'cooperate vigorously' strategy-level and k-1 'submit with self-reliance' strategy-levels is an asymmetric Nash-Simon equilibrium of* $G_{\mathcal{F}}$*.*

**Theorem 2.** *Let* $\tilde{V} \simeq \mathfrak{C}_{1,q}^{0}$*, with* $1 \leq q \leq 4$*. A k-tuple of 'defect' strategy-levels is a symmetric Nash-Simon equilibrium of* $G_{\mathcal{F}}$*. So is a k-tuple of 'cooperate vigorously' strategy-levels.*

**Theorem 3.** *Let* $\tilde{V}_{\mathcal{F}} \simeq \mathfrak{C}_{1,5}^{0}$*. A (symmetric) Nash-Simon equilibrium of* $G_{\mathcal{F}}$*, with* $\tilde{V}_{\mathcal{F}} \simeq \mathfrak{C}_{1,4}^{0}$*, is a (symmetric) Nash-Simon equilibrium of* $G_{\mathcal{F}}$*.*

Let us designate a Nash-Simon equilibrium in theorems 1 and 2 above, as *type*-0, if it is a $k$-tuple of 'defect' strategy-levels, and, otherwise, as *type*-I and *type*-II, respectively.

**Corollary 1.** *The dimensionality of a type-0 Nash-Simon equilibrium of* $G_{\mathcal{F}}$ *is* $0$*.*

**Corollary 2.** *Let* $\tilde{V} \simeq \mathfrak{C}_{0,q}^{0}$*. The dimensionality of a type-I Nash-Simon equilibrium of* $G_{\mathcal{F}}$ *is* $1$*.*

**Corollary 3.** *Let* $\tilde{V} \simeq \mathfrak{C}_{1,q}^{0}$*, with* $1 \leq q \leq 4$*. The dimensionality of a type-II Nash-Simon equilibrium of* $G_{\mathcal{F}}$ *is* $k$*.*

**Corollary 4.** *Let* $\tilde{V} \simeq \mathfrak{C}_{1,5}^{0}$*. The dimensionality of a type-II Nash-Simon equilibrium of* $G_{\mathcal{F}}$ *is* $16$*.*

**IV.5 Axiom.** *A dilemma game,* $G_{\mathcal{F}}$*, on the flow network* $\mathcal{F}$ *underlying* $\mathscr{O}$ *occurs in reduced form.*

By virtue of II.2.2 and II.3.2

**Proposition.** *Let* $\mathscr{O}$ *be super(a)symmetric. A type-II (resp., type-I) Nash-Simon equilibrium of the dilemma game,* $G_{\mathcal{F}}$*, underlying* $\mathscr{O}$ *respects the bistability (coexistence of two states) characterizing* $\mathscr{O}$ *under correlated variability. A type-0 Nash-Simon equilibrium of the same game respects the differential correlations characterizing* $\mathscr{O}$ *under correlated variability.*

REMARK 1: Bistability in one-dimensional, self-organizing neuronal circuits under correlated variability is discussed in [T.34], and a similar pattern of activity has been reported in two dimensions [T.35]. Differential correlations are discussed in [T.36]. Such correlations tend to be buried under more 'efficient' forms of correlations. Near-zero mean correlations have been experimentally observed in [T.37], their possible functional role being to prevent small correlations from accumulating and dominating network activity.

REMARK 2: Bistability associated with *slow* oscillations (observed during slow-wave sleep) in one-dimensional, self-organizing neuronal circuits has been discussed in [T.38] following the experimental observations in [T.39]. The hypothesis that *neural mirror matter* and associated structures [T.17] (Main Text/Fig. 4(c)-(d)) can explain cortical rhythmic activity during sleep does not contradict the observations in III.

**IV.6 Proposition 1.** *By functional duality, one obtains the co-dilemma game,* $\mathcal{G}_{\mathcal{F}}^{\perp}$ *(that is, the dilemma game* $\mathcal{G}_{\mathcal{F}^{\perp}}$ *), strategies in* $\mathcal{G}_{\mathcal{F}}$ *being replaced with co-strategies in* $\mathcal{G}_{\mathcal{F}}^{\perp}$ *. Statements structurally dual to those in* IV.4-IV.5 *hold true for the co-dilemma game* $\mathcal{G}_{\mathcal{F}}^{\perp}$ *.*

**Proposition 2.** *Assume that a dilemma game,* $\mathcal{G}_{\mathcal{F}}$ *, on the flow network* $\mathcal{F}$ *and a co-dilemma game,* $\mathcal{G}_{\mathcal{F}}^{\perp}$ *, on the co-flow network* $\mathcal{F}^{\perp}$ *occur. Then:*

$$\tilde{\mathcal{G}}_{\mathcal{F}} = \tilde{\mathcal{G}}_{\mathcal{F}}^{\perp} \ .$$

*That is, the reduced dilemma game,* $\tilde{\mathcal{G}}_{\mathcal{F}}$ *, of* $\mathcal{G}_{\mathcal{F}}$ *is self-dual to the reduced co-dilemma game,* $\tilde{\mathcal{G}}_{\mathcal{F}}^{\perp}$ *, of* $\mathcal{G}_{\mathcal{F}}^{\perp}$ *, under mirror super(a)symmetry.*

**IV.7** DEFINITION (**latent substate/neural representation**) We say that, at a particular instant in time, $t$, a neural subpopulation $c \in \mathcal{C}$ in $\mathcal{O}$ is encoding or representing a *latent substate* if there is a function $\boldsymbol{\rho}_t : \mathcal{C} \to \{0,1\}$ such that $\boldsymbol{\rho}_t(c) = 1$ .

We define $\mathrm{Rep}_t(\mathcal{O}) := (\boldsymbol{\rho}_t(1), \boldsymbol{\rho}_t(2), ..., \boldsymbol{\rho}_t(k))$ to be the $k$-tuple of *neural representations* of a latent state of $\mathcal{O}$ at $t$.

The $k$-tuple, $\mathrm{Rep}_t^{\perp}(\mathcal{O})$, of *neural co-representations* of a co-latent state of $\mathcal{O}$ is defined in an analogous way.

DEFINITION (**geometric dimensionality**) The instantaneous *geometric* or *Nash-Simon dimensionality* of neural representations of a latent state of $\mathcal{O}$ is defined as

$$d_t := \sum_c \boldsymbol{\rho}_t(c) = \mathscr{I}(\tilde{\boldsymbol{s}}^*),$$

and is identified with the dimensionality of some Nash-Simon equilibrium, $\tilde{\boldsymbol{s}}^*$, of the dilemma game, $\mathcal{G}_{\mathcal{F}}$, underlying $\mathcal{O}$ at $t$.

Since rotational or Lorentzian co-dynamics may also be observed in $\mathcal{O}$, the functionally dual notion of dimensionality may be similarly introduced, and the *complete* Nash-Simon dimensionality of neural (co-)representations of a (co-)latent state of $\mathcal{O}$, at $t$, may then be obtained as the sum $d_t + d_t^{\perp}$ .

**IV.8** DEFINITION (**Grothendieck game**) Let $S_{\mathcal{F}} := \left(\mathbf{M}, \mathbf{N}, \nu\right)$ be a dilemma game on $\mathcal{F}$, with $\mathbf{M} := \left(\mathrm{M}_c\right)_{\mathcal{C}}$ and $\mathbf{N} := \left(\left(\mathrm{N}_{cj}\right)_{j \in \mathcal{J}_c}\right)_{\mathcal{C}}$, such that $\left(\mathrm{M}_c, \mathrm{N}_{cj}\right)$ corresponds to a pair of elements in the Grothendieck ring, $\mathrm{K}\left(\widetilde{\mathrm{V}}\right)$, of $\widetilde{\mathrm{V}}$, with $\mathrm{N}_{cj} \subseteq \mathrm{M}_c \otimes_{\mathbb{R}} \left(\bigoplus_{i=0,1} \left(\bigoplus_{\mathcal{C}_i} \mathrm{M}_c\right)\right)$. Then $S_{\mathcal{F}}$ is a *spinorial* or *Grothendieck game* on $\mathcal{F}$.

We denote by $\overline{\mathrm{E}}_c := \left(\mathrm{E}_{cj}\right)_{j \in \mathcal{J}_c}$ the indexed family of all possible $c$-edge modules on $\mathcal{V}$, subject to Axiom II.1.2. The $k$-tuple of canonical injections $\rho : \left(\mathcal{V}_c\right)_{\mathcal{C}} \to \left(\mathrm{V}_c\right)_{\mathcal{C}}$ defines a structure-preserving map $\left(\left(\mathcal{V}_c\right)_{\mathcal{C}}, \left(\overline{\mathcal{E}}_c\right)_{\mathcal{C}}, \left(\varepsilon_c\right)_{\mathcal{C}}\right) \to \left(\left(\mathrm{V}_c\right)_{\mathcal{C}}, \left(\overline{\mathrm{E}}_c\right)_{\mathcal{C}}, \left(\boldsymbol{\epsilon}_c\right)_{\mathcal{C}}\right)$ between dilemma games on $\mathcal{F}$. Let $\mathbf{V} := \left(\mathrm{V}_c\right)_{\mathcal{C}}$, $\mathbf{E} := \left(\overline{\mathrm{E}}_c\right)_{\mathcal{C}}$, and $\boldsymbol{\epsilon} := \left(\boldsymbol{\epsilon}_c\right)_{\mathcal{C}}$. The *spinorial completion* of $\mathcal{G}_{\mathcal{F}}$ is the spinorial game $\widetilde{S}_{\mathcal{F}} := \left(\mathbf{V}, \mathbf{E}, \boldsymbol{\epsilon}\right)$, together with the $\mathcal{F}$-game monomorphism $\rho : \mathcal{G}_{\mathcal{F}} \to \widetilde{S}_{\mathcal{F}}$.

By universality of free modules, for every $\mathcal{F}$-game homomorphism, $\psi : \mathcal{G}_{\mathcal{F}} \to S_{\mathcal{F}}$, from $\mathcal{G}_{\mathcal{F}}$ to a spinorial game $S_{\mathcal{F}}$, there is a unique homomorphism, $\overline{\psi} : \widetilde{S}_{\mathcal{F}} \to S_{\mathcal{F}}$, of spinorial games on $\mathcal{F}$, such that $\overline{\psi} \circ \rho = \psi$.

**IV.9** It is of course necessary to comment on how the present report accounts for the functional variation observed across individual nervous systems [T.40], and species. We hold that, under Axiom IV.3:

**Proposition.** *Nash-Simon equilibria of universal spinorial games respect Edelman's neural Darwinism hypothesis.*

**IV.10** Further references: [T.41-46].